\documentclass[sigconf,screen]{acmart}
\usepackage{times}
\usepackage{subfig}
\usepackage{wrapfig}
\usepackage{float}
\usepackage{multirow, graphicx}
\usepackage{algorithm}
\usepackage{algpseudocode} %algorithm: use algorithmicx package, to refer to http://en.wikibooks.org/wiki/LaTeX/Algorithms_and_Pseudocode#Typesetting_using_the_algorithmicx_package
\usepackage{amsmath}

\usepackage{xspace}

\usepackage{url}

\usepackage{breakurl}

\usepackage{sty/widetext} %show equation in two columns

\usepackage{stackrel}

\ifdefined\TTSTYLETARGETreport
\usepackage[
    backend=bibtex,
    maxbibnames=100,
    firstinits=true,
    bibstyle=numeric,
    citestyle=numeric
]{biblatex}
\fi

\usepackage{pifont}% http://ctan.org/pkg/pifont
\newcommand{\cmark}{\ding{51}}%
\newcommand{\xmark}{\ding{55}}%
\newcommand{\ignore}[1]{}

\usepackage{bbding} 

\usepackage{todonotes}
\usepackage{graphicx, multirow}

\providecommand{\defeq}{\stackrel{\text{def}}{=}}
\usepackage{tabularx, multirow, rotating} %tables, dependent on multirow package.
\usepackage{subfig} %figures
\usepackage{algorithm} %algorithm
\usepackage{algpseudocode} %algorithm: use algorithmicx package, to refer to http://en.wikibooks.org/wiki/LaTeX/Algorithms_and_Pseudocode#Typesetting_using_the_algorithmicx_package
\usepackage{ulem} % strike out text
\usepackage{color} % colored text

\usepackage{listings} 
\lstdefinestyle{Oracle}{basicstyle=\ttfamily,
                        keywordstyle=\lstuppercase,
                        emphstyle=\itshape,
                        showstringspaces=true,
                        }
\makeatletter
\newcommand{\lstuppercase}{\uppercase\expandafter{\expandafter\lst@token
                           \expandafter{\the\lst@token}}}
\newcommand{\lstlowercase}{\lowercase\expandafter{\expandafter\lst@token
                           \expandafter{\the\lst@token}}}
\makeatother

\usepackage{tikz}

\newcommand{\ballnumber}[1]{\tikz[baseline=(myanchor.base)] \node[circle,fill=.,inner sep=1pt] (myanchor) {\color{-.}\bfseries\scriptsize #1};}
\newif\ifboldnumber

\algrenewcommand\alglinenumber[1]{%
  \footnotesize\ifboldnumber\bfseries\fi\global\boldnumberfalse#1:}

\usepackage[misc]{ifsym}

\AtBeginDocument{%
  \providecommand\BibTeX{{%
    \normalfont B\kern-0.5em{\scshape i\kern-0.25em b}\kern-0.8em\TeX}}}

\acmConference[Extended from ESEC/FSE '21]{Proceedings of the 29th ACM Joint European Software Engineering Conference and Symposium on the Foundations of Software Engineering}{August 23--28, 2021}{Athens, Greece}
\begin{document}

\providecommand{\sysname}{\textsc{iBatch}\xspace}
%\title{Towards Practical and Cost-Effective Batching of Smart-Contract Invocations on Ethereum}
\title{\sysname: Saving Ethereum Fees via Secure and Cost-Effective Batching of Smart-Contract Invocations}
%\title{Smart Contract Execution at Reduced Fees via Secure and Cost-Effective Batching on Ethereum}

\subtitle{Extended version from the ESEC/FSE'21 paper}

\author{Yibo Wang}
\email{ywang349@syr.edu}
\affiliation{%
  \institution{Syracuse University}
  \city{Syracuse}
  \state{NY}
  \country{USA}
%  \postcode{13244}
}

\author{Kai Li}
\email{kli111@syr.edu}
\affiliation{%
  \institution{Syracuse University}
  \city{Syracuse}
  \state{NY}
  \country{USA}
%  \postcode{13244}
}

\renewcommand{\footnotemark}{{\Envelope}}
\author{Yuzhe Tang}
\authornote{\Envelope\ \xspace{}Yuzhe Tang is the corresponding author.}

\email{ytang100@syr.edu}
\affiliation{%
  \institution{Syracuse University}
  \city{Syracuse}
  \state{NY}
  \country{USA}
%  \postcode{13244}
}

\author{Jiaqi Chen}
\email{jchen217@syr.edu}
\affiliation{%
  \institution{Syracuse University}
  \city{Syracuse}
  \state{NY}
  \country{USA}
%  \postcode{13244}
}

\author{Qi Zhang}
\email{qzhang71@syr.edu}
\affiliation{%
  \institution{Syracuse University}
  \city{Syracuse}
  \state{NY}
  \country{USA}
%  \postcode{13244}
}

\author{Xiapu Luo}
\email{csxluo@comp.polyu.edu.hk}
\affiliation{%
  \institution{\small Hong Kong Polytechnic University}
  \city{Kowloon}
  \state{Hong Kong}
  \country{China}
%  \postcode{SAR}
}

\author{Ting Chen}
\email{brokendragon@uestc.edu.cn}
\affiliation{%
  \institution{\small University of Electronic Science and Technology of China}
  \city{Chengdu}
  \state{Sichuang}
  \country{China}
%  \postcode{}
}

%\email{webmaster@marysville-ohio.com}
%\affiliation{%
%  \institution{Institute for Clarity in Documentation}
%  \streetaddress{P.O. Box 1212}
%  \country{USA}

%\author{Lars Th{\o}rv{\"a}ld}
%\affiliation{%
%  \institution{The Th{\o}rv{\"a}ld Group}
%  \streetaddress{1 Th{\o}rv{\"a}ld Circle}
%\email{larst@affiliation.org}

\renewcommand{\shortauthors}{Yibo Wang, Qi Zhang, Kai Li, Yuzhe Tang, Jiaqi Chen, Xiapu Luo, and Ting Chen}

\begin{abstract}
This paper presents \sysname, a middleware system running on top of an operational Ethereum network to enable secure batching of smart-contract invocations against an untrusted relay server off-chain. \sysname does so at a low overhead by validating the server's batched invocations in smart contracts without additional states. 
The \sysname mechanism supports a variety of policies, ranging from conservative to aggressive batching, and can be configured adaptively to the current workloads.
\sysname automatically rewrites smart contracts to integrate with legacy applications and support large-scale deployment.

We built an evaluation platform for fast and cost-accurate transaction replaying and constructed real transaction benchmarks on popular Ethereum applications. With a functional prototype of \sysname, we conduct extensive cost evaluations, which shows \sysname saves $14.6\%\sim{}59.1\%$ Gas cost per invocation with a moderate 2-minute delay and $19.06\%\sim{}31.52\%$ Ether cost per invocation with a delay of $0.26\sim{}1.66$ blocks. 
\end{abstract}

\ignore{
This paper presents iBatch, a middleware system running on top of an operational Ethereum network to enable secure batching of smart-contract invocations against an untrusted relay server off-chain. iBatch does so at a low overhead by validating the server's batched invocations in smart contracts without additional states. 
The iBatch mechanism supports a variety of policies, ranging from conservative to aggressive batching, and can be configured adaptively to the current workloads.
iBatch automatically rewrites smart contracts to integrate with legacy applications and support large-scale deployment.

We built an evaluation platform for fast and cost-accurate transaction replaying and constructed real transaction benchmarks on popular Ethereum applications. With a functional prototype of iBatch, we conduct extensive cost evaluations, which shows iBatch saves 14.6%-59.1% Gas cost per invocation with a moderate 2-minute delay and 19.06%-31.52% Ether cost per invocation with a delay of 0.26-1.66 blocks. 
}

%\begin{CCSXML}
%<ccs2012>
%  <concept_id>10010520.10010553.10010562</concept_id>
%  <concept_desc>Computer systems organization~Embedded systems</concept_desc>
%  <concept_significance>500</concept_significance>
%  <concept_id>10010520.10010575.10010755</concept_id>
%  <concept_desc>Computer systems organization~Redundancy</concept_desc>
%  <concept_significance>300</concept_significance>
%  <concept_id>10010520.10010553.10010554</concept_id>
%  <concept_desc>Computer systems organization~Robotics</concept_desc>
%  <concept_significance>100</concept_significance>
%  <concept_id>10003033.10003083.10003095</concept_id>
%  <concept_desc>Networks~Network reliability</concept_desc>
%  <concept_significance>100</concept_significance>
%</ccs2012>
%\end{CCSXML}
%
%\ccsdesc[500]{Computer systems organization~Embedded systems}

\keywords{Blockchains, smart contracts, DeFi, cost effectiveness, replay attacks}

\maketitle

\providecommand{\ssssp}{{\sc SS\_SSP}\xspace}
\newcommand{\tremark}[1]{\footnote{\textcolor{red}{(Ting's comment: #1)}}}
\newcommand{\xremark}[1]{\footnote{\textcolor{red}{(Xin's comment: #1)}}}
\newcommand{\jj}[1]{\footnote{\textcolor{blue}{(Jiyong: #1)}}}
\newcommand{\yz}[1]{\footnote{\textcolor{red}{(Yuzhe: #1)}}}

\definecolor{mygreen}{rgb}{0,0.6,0}
\lstset{ %
  backgroundcolor=\color{white},   % choose the background color; you must add \usepackage{color} or \usepackage{xcolor}
  %basicstyle=\footnotesize\ttfamily,        % the size of the fonts that are used for the code
  basicstyle=\scriptsize\ttfamily,        % the size of the fonts that are used for the code
  breakatwhitespace=false,         % sets if automatic breaks should only happen at whitespace
  breaklines=true,                 % sets automatic line breaking
  captionpos=b,                    % sets the caption-position to bottom
  commentstyle=\color{mygreen},    % comment style
  deletekeywords={...},            % if you want to delete keywords from the given language
  escapeinside={\%*}{*)},          % if you want to add LaTeX within your code
  extendedchars=true,              % lets you use non-ASCII characters; for 8-bits encodings only, does not work with UTF-8
  %frame=single,                    % adds a frame around the code
  keepspaces=true,                 % keeps spaces in text, useful for keeping indentation of code (possibly needs columns=flexible)
  keywordstyle=\color{blue},       % keyword style
  language=Java,                 % the language of the code
  morekeywords={*,...},            % if you want to add more keywords to the set
  numbers=left,                    % where to put the line-numbers; possible values are (none, left, right)
  numbersep=5pt,                   % how far the line-numbers are from the code
  numberstyle=\scriptsize\color{black}, % the style that is used for the line-numbers
  rulecolor=\color{black},         % if not set, the frame-color may be changed on line-breaks within not-black text (e.g. comments (green here))
  showspaces=false,                % show spaces everywhere adding particular underscores; it overrides 'showstringspaces'
  showstringspaces=false,          % underline spaces within strings only
  showtabs=false,                  % show tabs within strings adding particular underscores
  stepnumber=1,                    % the step between two line-numbers. If it's 1, each line will be numbered
  stringstyle=\color{mymauve},     % string literal style
  tabsize=2,                       % sets default tabsize to 2 spaces
  title=\lstname,                  % show the filename of files included with \lstinputlisting; also try caption instead of title
  moredelim=[is][\bf]{*}{*},
}

\newcommand{\tangSide}[1]{\todo[caption={},color=cyan!20!]{{\scriptsize #1}}}
\setlength{\marginparwidth}{1.5cm}

\section{Introduction}
The recent paradigm shift to building decentralized applications (DApps) on blockchains has nurtured a number of fast-growing domains, such as decentralized finance (DeFi), decentralized online gaming, et al. that have the potential of disrupting conventional business in finance, gaming, et al. The core value brought by DApps is their decentralized system architecture that is amendable to tackle the mistrust crisis in many security-oriented businesses (e.g., ``trusted'' authorities are constantly caught misbehaving).
However, despite the attractive trustless architecture and moderate popularity in practice, an impediment to DApps' broader adoption is their intensive use of underlying blockchain and the associated high costs.
Ethereum~\cite{me:eth}, the second largest blockchain after Bitcoin and the most popular DApp platform, charges a high unit cost for data movement (via transactions) and for data processing (via smart-contract execution). For instance, sending one-megabyte application data to Ethereum costs $17.5$ Ether or more than $25,000$ USD (at the exchange rate as of Jan. 2021), which is much more expensive than alternative centralized solutions (e.g., cloud services) and has scared away customers (e.g., Binance~\cite{me:bsc:flee}).

Towards cost-effective use of blockchains, existing research mainly tackles the problem from the angle of designing new protocols at blockchain layer one (i.e., redesigning the consensus protocol and building a new blockchain system~\cite{DBLP:conf/sp/Kokoris-KogiasJ18,DBLP:conf/nsdi/EyalGSR16}) and at layer two (i.e., by offloading the workload from the blockchain to off-chain clients, such as in payment channel networks~\cite{me:lightning,DBLP:journals/corr/MillerBKM17,DBLP:journals/corr/abs-1804-05141,DBLP:conf/ccs/DziembowskiFH18}). However, these new protocols are designed without the legacy platform of an operational blockchain and deployed DApps in mind and result in unsatisfactory deployability: 
For instance, existing protocols either require bootstrapping a brand new blockchain network (as in the layer-one approach) or develop from scratch the on-chain and off-chain components of a DApp (i.e., to support payment networks). As a result, there is a lack of adoption of these protocols among legacy DApps at scale.

This work aims at optimizing blockchain costs among legacy DApps on Ethereum.
Towards the goal, we focus on designing a middleware system running on top of unmodified Ethereum platform and DApp clients. We also develop software tools to facilitate integrating the middleware with legacy DApp clients and smart contracts.

To motivate our approach, consider a typical DApp architecture where a DApp client holding an Ethereum account sends a transaction on the Ethereum blockchain to invoke a smart-contract function there. A typical DApp's smart contract runs event-driven logic, and a popular DApp would receive a ``large'' number of ``small'' invocations: 1) An individual invocation is often with a small amount of data and triggers few lines of smart-contract code; think as an example the \texttt{transfer()} function in an ERC20 token smart contract. 2) A popular DApp features an intensive stream of invocations that arrive at a high rate. 
{
%\color{red}~\tangSide{Response to Q6' [\#258D.i]}
This workload characteristic holds over time, as we verified on various Ethereum traces (see the IDEX trace in \S~\ref{sec:motives} and Chainlink/BNB/Tether traces in \S~\ref{sec:eval}), and it is also corroborated by external Ethereum exploration services~\cite{me:etherscan:txchart,me:bitinfo:txchart}.
}
The workload with a high rate of small invocations renders the transaction fee a significant cost component that alone is worth optimization.
To optimize the transaction fee, a natural idea is to batch multiple smart-contract invocations in a single transaction so that the fee can be amortized~\cite{me:bitcoinbatching,DBLP:conf/esorics/FrowisB19}. For instance, under Ethereum's current block limit, one can theoretically batch up to $20$ normal invocations in a transaction, leading to a potential fee reduction by $\frac{1}{20}\times$. By this promise, invocation batching has long been craved for among Ethereum developers, evidenced by a number of Ethereum Improvement Proposals (EIPs)~\cite{me:eip2711,me:eip3005,me:eip:7525}. 
{
%\color{red}~\tangSide{Response to Q2 [\#258A]}
Despite the strong interest, it still lacks real-world support of invocation batching in Ethereum, as these EIPs are not made into production after years of discussion. We believe this unsatisfactory status is due to the design challenges raised by the tradeoff among batching's security, cost-effectiveness and timeliness (short delay), as presented next.
}

%Despite the strong interest, there is a lack of real-world support of invocation batching in the Ethereum ecosystem; for instance, all the EIPs mentioned above are in the draft status, even after years of discussion and development. To support invocation batching on Ethereum with actual cost saving, there are technical challenges down the road.

\begin{figure}[!bthp]
\centering
\includegraphics[width=0.5\textwidth]{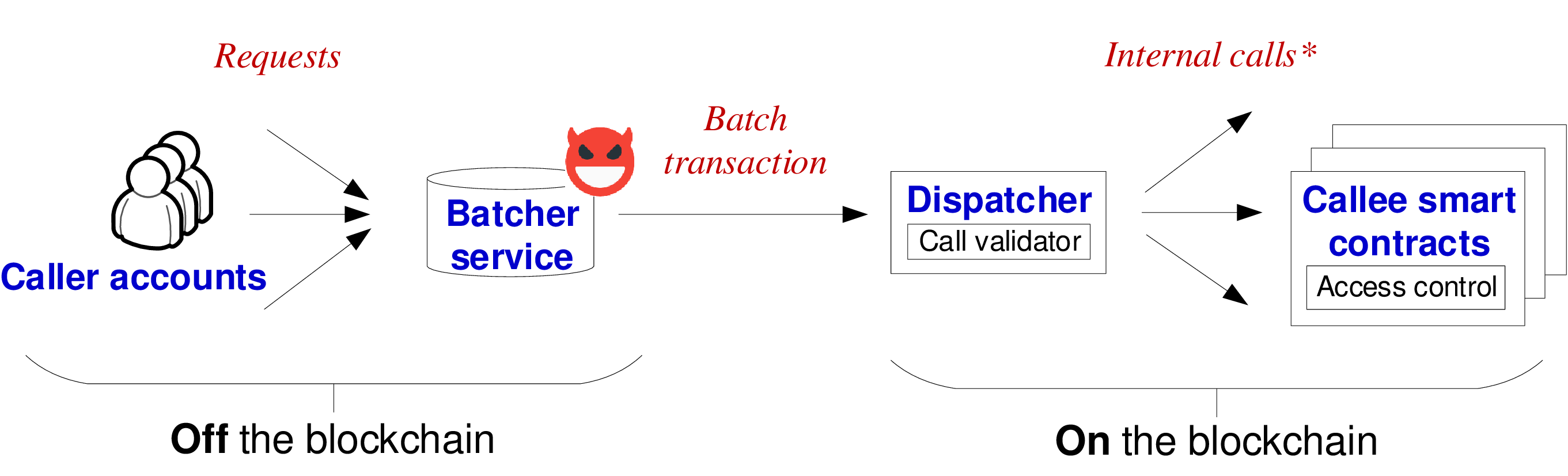}
\caption{Batching smart-contract invocations in Ethereum: {\normalfont Note that the \texttt{Dispatcher} can be a standalone smart contract or be a function inlining the callee function; in the latter case, the ``internal call$^*$'' is straight-line code execution.}}
\label{fig:ideabatching}
\end{figure}

%\vspace{3pt}
\noindent\textbf{
Challenges}: 
First, Ethereum does not have native support of batching in the sense that an Ethereum transaction transfers Ether from one account to another account. This is different from Bitcoin and other blockchains whose transaction can encode multiple coin transfers. This difference renders the existing architectures to batch transfers in non-Ethereum blockchains~\cite{me:bitcoinbatching,DBLP:conf/fc/BonneauNMCKF14,DBLP:conf/ndss/HeilmanABSG17,DBLP:journals/popets/MeiklejohnM18,DBLP:journals/iacr/SeresNBB19} inapplicable to batch invocations on Ethereum. To address the challenge, we introduce two intermediaries between a caller account and a callee smart contract. As depicted in Figure~\ref{fig:ideabatching}, they are a relay service off-chain, called \texttt{Batcher}, and an on-chain component, called \texttt{Dispatcher}. The \texttt{Batcher}'s job is to batch multiple invocation requests sent from the caller accounts and send them in a single transaction to the \texttt{Dispatcher}, which further unmarshalls the original invocations and relays them individual to the intended callee smart contracts. 

Second, the off-chain \texttt{Batcher} service need not be trusted by the callers (who, in a decentralized world, are reluctant to trust any third-parties beyond the blockchain). Defending against the untrusted \texttt{Batcher} incurs overhead that may offset the cost saving from batching and instead result in net cost increases. Specifically, in our threat model, the adversarial \texttt{Batcher} is financially incentivized to mount attacks and to modify, forge, replay or omit the invocation requests in the batch transaction; for instance, replaying a \texttt{transfer()} of an ERC20 token can benefit the receiver of the transfer. To defend against the threat, a baseline design is to run the entire transaction validation logic in the trusted \texttt{Dispatcher} smart contract on-chain, which bloats the contract program and incurs overhead (e.g., to maintain additional program states). Our evaluation study 
(in Figure~\ref{fig:batchsize:1} in \S~\ref{appdx:sec:eval:micro}
%in Technical Report~\cite{me:ibatch:tr}
shows this baseline denoted by B2 increases the net cost per invocation rather than decreasing it.
For secure and cost-effective batching, we propose a security protocol that allows off-chain DApp callers in the same batch to jointly sign the batch transaction so that the additional program states (e.g., the per-account nonces as a defense to replaying attacks) can be offloaded offline and the \texttt{Dispatcher} smart contract can be {\it stateless}, rendering overhead low and leading to positive net cost savings.

%entails smart-contract code to verify signatures and check on nonces (a nonce is a monotonic counter per account and is used to mitigate transaction relay attacks). This baseline design, named B2 in \S~\ref{sec:baseline:b2}, incurs updating smart-contract storage per invocation and introduces so high overhead that actually offsets the saving of transaction fee by batching, as verified in our evaluation in \S~\ref{sec:eval:micro}. 

Third, batching requires to wait for enough invocations and can introduce delay to when the batched invocations are included in the blockchain. 
For the many DApps sensitive to invocation timing (e.g., real-time trading, auctions and other DeFi applications), such delay is undesirable. 
To attain delay-free batching, we propose to use the transaction price to the rescue. Briefly, Ethereum blockchain admits a limited number of transaction per block and prioritizes the processing of incoming transactions with a higher ``price'' (i.e., the so-call Gas price which is the amount of Ether per each computation unit paid to miners). Thus, our idea is to generate a batch transaction with a higher price so that it can be included in the blockchain more quickly, and this saved time can offset the waiting time caused by batching, resulting in an overall zero delay in blockchain inclusion. We propose an online mechanism to conservatively batch invocations originally in one block and carefully set Gas price of batch transactions with several heuristics to counter the limited knowledge in online batching.   

\vspace{3pt}\noindent\textbf{
Systems solutions}: 
Overall, this work systematically addresses the challenges above and presents a comprehensive framework, named \sysname, that incorporates the proposed techniques under one roof. 
\sysname{} includes the middleware system of \texttt{Dispatcher} and \texttt{Batcher} and a series of policies that configure the system to adapt the batching to specific DApps' workloads.
Concretely, the middleware system exposes knobs to tune the batching in timing (how long to wait for invocations to be batched), target invocations (what invocations to batch) and other conditions. Through this, policies that range from conservative to aggressive batching are proposed, so that the system can be tailored to the different needs of DApps. 
For instance, the DApps sensitive to invocation timing can be best supported by the conservative batching policy with minimal delay. Other DApps more tolerable with delays can be supported by more aggressive batching policies which result in higher degree of cost saving. We demonstrate the feasibility of \sysname's middleware design by building a functional prototype with Ethereum's Geth client~\cite{me:geth}. Particularly, we statically instrument Geth to hook the \texttt{Batcher}'s code. 
% (note that only the Geth co-residing the \texttt{Batcher} needs to be modified and other Geth nodes in the Ethereum network need not).

{
%\color{red}
We further address the integration of \sysname{} with legacy DApps and the operational Ethereum network by automatically rewriting their smart contracts. Briefly, with batching, the internal calls are sent from \texttt{Dispatcher} (instead of the original caller account), which makes them unauthorized access to the original callee, leading to failed invocations.
In \sysname, we propose techniques to rewrite callee smart contracts, particularly their access-control structure to white-list \texttt{Dispatcher}.
The proposed bytecode rewriter will be essential to support the majority of legacy smart contracts deployed on Ethereum mainnet without Solidity source code.
%~\tangSide{Response to Q4' [\#258D.iii]}
We acknowledge the recent Ethereum development EIP-3074~\cite{me:eip:3074}, which, if made into an operational Ethereum network, will facilitate \sysname's integration without rewriting smart contracts (details in \S~\ref{sec:eip3074}).
}

\vspace{3pt}\noindent\textbf{
Systematic evaluation}:
We systematically evaluate the invocation cost and delay in \sysname, under both real and synthetic workloads. First, we build a fast transaction-replay engine that executes transactions at a much higher speed than the transactions are originally included in the blockchain. This allows us to conduct large-scale measurements, say replaying a trace of transactions that last for months in real life within hours in the experiments. Second, we collect the trace of transactions/calls under four representative DApps, that is, IDEX~\cite{me:idex} representing decentralized exchanges (DEX), BNB~\cite{me:binance:contract} and Tether~\cite{me:tether} for tokens, and Chainlink~\cite{me:chainlink} for data feeding. From there, we build a benchmark of traces that can be replayed in our platform. Third, we conduct extensive evaluations based on the developed platform (i.e., replay engine, benchmarks and prototype we built). The target performance metrics are the system's costs (in terms of Ether and Gas) and delays between invocation submission time and block inclusion time. 

The result under the BNB-token/IDEX/Chainlink trace shows that \sysname{} configured with a time window of $120$ seconds to batch all invocations can save around $50\%/24\%/17.6\%$ of the Gas per invocation of the unbatched baseline. 
For delay-sensitive DApps, as we evaluate under the workloads of Tether tokens, 
\sysname{} can save $19.06\%$ ($31.52\%$) cost at the expense of causing a delay of $0.26$ ($1.66$) blocks.

{
%\color{red}
\vspace{3pt}\noindent\textbf{
Contributions}:
This work makes the following contributions:
%~\tangSide{Response to Q3' [\#258D.ii]}

\vspace{2pt}\noindent$\bullet$\textit{
Security protocol}: We design a lightweight security protocol for batching of smart contract invocations in Ethereum without trusting third-party servers (i.e., the \texttt{Batcher}). The security protocol defends against a variety of invocation manipulations. New techniques are proposed to jointly sign invocations off-chain and validate invocations on-chain without states against replay attacks.

\vspace{2pt}\noindent$\bullet$\textit{
Cost-effective systems}: We design a middleware system implementing the above protocol and propose batching policies from conservative to aggressive batching. Particularly, we propose an online mechanism to optimize the cost without delaying invocation execution. We further address the integration with the current Ethereum client by automatically rewrite smart contracts.

\vspace{2pt}\noindent$\bullet$\textit{
Systematic evaluation}: We built an evaluation platform for fast and cost-accurate transaction replaying and constructed transaction benchmarks on popular Ethereum applications. With a functional prototype of \sysname, we conduct extensive cost evaluations, which shows \sysname saves $14.6\%\sim{}59.1\%$ Gas cost per invocation with a moderate 2-minute delay and $19.06\%\sim{}31.52\%$ Ether cost per invocation with a delay of $0.26\sim{}1.66$ blocks.

Overall, this work tackles the design tradeoff among security, cost and delays by batching invocations. While implementation and evaluation are on Ethereum, we believe the design tradeoffs  and choices in this work are generically applicable to smart-contract platforms beyond Ethereum.
}

\vspace{3pt}\noindent\textbf{
Roadmap}: Section \S~\ref{sec:research} formulates the research. 
\S~\ref{sec:baseline} presents the \sysname's security protocol.
\sysname's batching policies are described in \S~\ref{sec:customization}. 
\S~\ref{sec:rewrite} presents the smart-contract rewriters to facilitate \sysname's integration with legacy smart contracts.
\S~\ref{sec:eval} shows the evaluation results in cost and invocation delay.
Related works are described in \S~\ref{sec:rw} and conclusion in \S~\ref{sec:conclude}. 

\vspace{-0.1in}
\section{Research Formulation}
\label{sec:research}

\subsection{Motivating Example}
\label{sec:motives}
\begin{figure*}[!bthp]
\centering
    \subfloat[The IDEX system model and protocol: 
The example scenario shows running IDEX among six Ethereum accounts: Three user accounts (\texttt{maker}, \texttt{taker} and \texttt{IDEX2}) and three contracts (\texttt{mToken}, \texttt{tToken} and \texttt{IDEX1}).]{%
\label{fig:idex}
\includegraphics[width=0.425\textwidth]{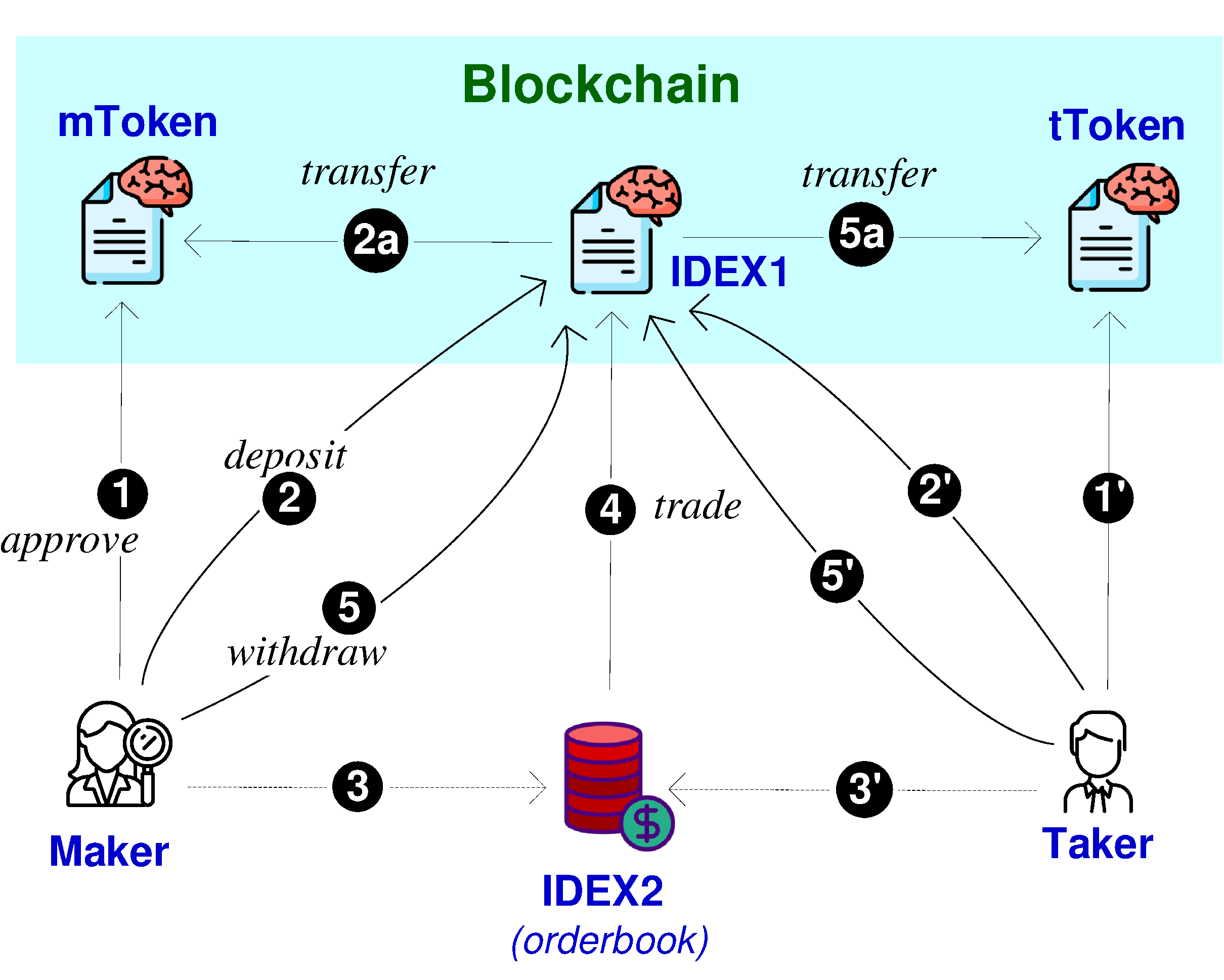}
}
\hspace{0.2in}
    \subfloat[Distribution of \texttt{trade} calls over Ethereum blocks; in this figure, $X=5,Y=11.4$ means in $11.4\%$ of Ethereum blocks, the number of \texttt{trade} calls per block is between $2$ and $5$.]{%
\label{fig:overblocks}
\includegraphics[width=0.4\textwidth]{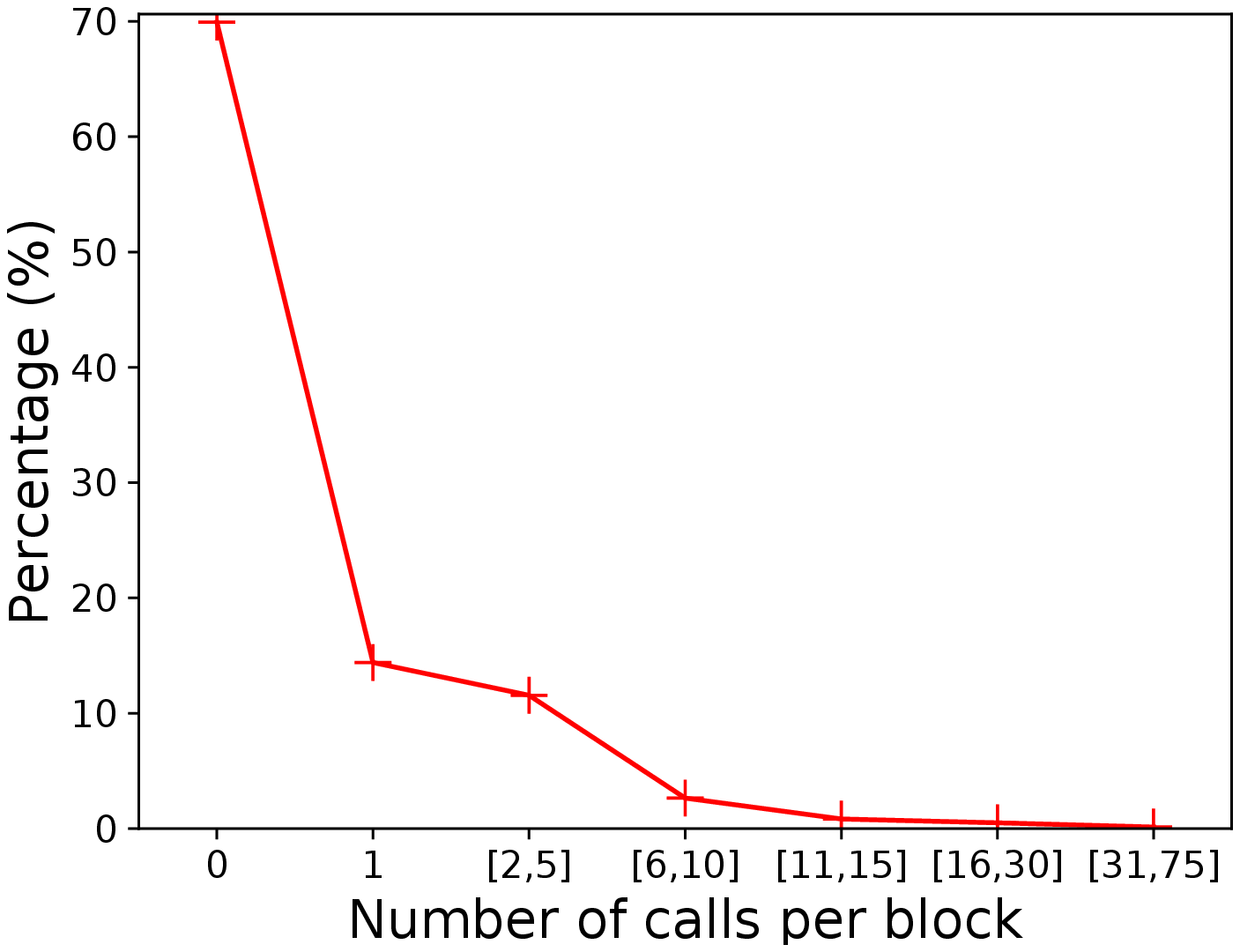}
}
\vspace{0.1in}
\caption[]{IDEX protocol and call distribution: 
The protocol execution in Figure~\ref{fig:idex} involves five steps: 1) The \texttt{maker} deposits her tokens to \texttt{IDEX1}, which invokes three functions: \ballnumber{1} \texttt{maker.approve()}, \ballnumber{2} \texttt{maker.deposit()} and \ballnumber{2a} \texttt{IDEX1.transferFrom(maker,DEX1)}. 
2) The taker similarly deposits his tokens by issuing \ballnumber{1'} \texttt{taker.approve()}, \ballnumber{2'} \texttt{taker.deposit()} and \texttt{IDEX1.transferFrom(taker,DEX1)} (not shown in the figure).
3) The maker and taker sends their respective selling and buying orders (\ballnumber{3} and \ballnumber{3'}) to the off-chain \texttt{IDEX2}, who match-make orders in an order-book. 4) The owner \texttt{IDEX2} calls contract \texttt{IDEX1}'s function \texttt{trade(taker,maker)} (\ballnumber{4}) to execute the trade on-chain. 5) The maker issues \texttt{withdraw} (\ballnumber{5}) which further sends \texttt{transfer()} calls (\ballnumber{5a}) to \texttt{tToken} contract. Similarly, the taker can submit calls to withdraw her tokens (\ballnumber{5'}).
}
\end{figure*}

We use a real-world scenario, namely IDEX~\cite{me:idex,me:idex:website}, to motivate our work. IDEX is a decentralized exchange protocol that allows owners of different ERC20 tokens to exchange their tokens at the preferred price/volume. 
Consider that account Alice sells her tokens \texttt{mToken} to another account Bob in return of his tokens \texttt{tToken}. To do so, Alice makes an order to be taken by Bob, and Alice (Bob) is called a \texttt{maker} (a \texttt{taker}). The IDEX protocol is executed among six Ethereum accounts include a \texttt{maker} account, a \texttt{taker} account, maker's token contract \texttt{mToken}, taker's token contract \texttt{tToken}, the core IDEX smart contract \texttt{IDEX1}~\cite{me:idex1}, and \texttt{IDEX1}'s off-chain owner \texttt{IDEX2}~\cite{me:idex2}. The protocol execution is depicted and described in Figure~\ref{fig:idex}. 

In particular, there are four types of transactions in IDEX that invoke smart contracts, that is, maker's (taker's) call to \texttt{approve} her (his) token contract (i.e., \ballnumber{1} and \ballnumber{1'}), maker's (taker's) call to \texttt{deposit} to \texttt{IDEX1} (i.e., \ballnumber{2} and \ballnumber{2'}), IDEX2's call to \texttt{trade} on \texttt{IDEX1} (i.e., \ballnumber{4}), and maker's (taker's) call to \texttt{withdraw} (i.e., \ballnumber{5}).
Among the transaction-triggered external calls, \texttt{trade} is most intensively invoked. As we examine the Ethereum history via Ethereum-ETL service on Google BigQuery~\cite{me:ethtrace:bigquery}, $61.59\%$ of the invocations received by the \texttt{IDEX1} contract 
{
%\color{black}~\tangSide{Response to Q6' [\#258D.i]}
from its launch on Sep. 27, 2017 to Feb. 23, 2019 are on \texttt{trade()}.\footnote{
%\color{black} 
We did not take the \texttt{IDEX} transactions after Feb. 2019 when IDEX's traffic started to decline and was then shadowed by other more popular DEX, such as Uniswap~\cite{me:uniswapv2}. Here, we stress that although our IDEX's trace ends in Feb. 2019 (as of this writing in May 2021), Ethereum's transaction rate steadily increases over time. Particularly, recent years see drastic rate growth as Ether price soars since early 2020. This is verified by the more recent traces we collected in 2020, such as Chainlink and Tether tokens, as in the cost evaluation in \S~\ref{sec:eval}, and also corroborates external Ethereum exploration websites~\cite{me:bitinfo:txchart,me:etherscan:txchart}.}}
More importantly, {\it the \texttt{trade} invocations are so intensively issued that many of them wind up in the same Ethereum block.} 
We measured the number of \texttt{trade} calls in the same Ethereum block, on the call trace above. 
Figure~\ref{fig:overblocks} plots the cumulative distribution of Ethereum blocks by the per-block call number. For instance, about $30\%$ of Ethereum blocks have more than one \texttt{trade} calls in them, $5\%$ of blocks have more than four \texttt{trade} calls, and $0.36\%$ blocks have 20 \texttt{trade} calls. 
If one batches the 20 \texttt{trade} invocations of these Ethereum blocks into a single transaction, the transaction fee can be reduced to $\frac{1}{20}$, although it may incur additional costs for smart-contract execution. 
In general, for blocks with $X$ \texttt{trade} calls, one can batch the calls into one transaction, leading to a $X$-fold fee reduction. By plugging into $X$ the measurement results in Figure~\ref{fig:overblocks}, we can expect the overall fee-saving in the case IDEX to be $10.7\%$. This is the saving from \texttt{trade} calls only. Note that because the original \texttt{trade} calls are in the same block, batching them in a single transaction does not introduce additional delay/inconsistency.
% 1) submitted by the same account, 2) to the same contract, and 3) included in the same blockchain block, 

Generally, there are four types of batching strategies: {\bf Type S1)} Batch invocations of the same caller and same callee, such as all \texttt{trade} calls from the same caller (\texttt{IDEX2}) and sent to the same callee smart contract (\texttt{IDEX1}),
{\bf S2)} batch invocations of different callers and the same callee, such as all the \texttt{deposit} calls, 
{\bf S3)} batch invocations of the same caller and different callees, and 
{\bf S4)} batch invocations of different callers and different callees, such as the \texttt{approve} calls in the case of IDEX.
We mainly consider the general case of S4 in the paper and will tailor the system to different invocation types in \S~\ref{sec:customization}.

{
%\color{blue}
\vspace{-0.1in}
\subsection{Threat Model}
\label{sec:systemmodel}
\label{sec:trustmodel}

Recall the system model in Figure~\ref{fig:ideabatching} that introduces
the \texttt{Batcher} and \texttt{Dispatcher}, as two intermediaries between caller accounts and callee smart contracts.
For generalizability, our threat model considers an untrusted third-party \texttt{Batcher}. For instance, in the case of IDEX, the \texttt{Batcher} can batch \texttt{approve}, \texttt{deposit} and \texttt{trade}, and does not require the trust from their callers. The third-party \texttt{Batcher} can mount attacks to forge, replay, modify and even omit the invocations from the callers. Our model assumes unmodified the trust relationship among callers; for instance, if there is a counterparty risk between a maker account and taker account in the vanilla IDEX, the same trust remains in \sysname.

The smart contracts, including both \texttt{Dispatcher} and application contracts, are trusted in terms of program security (no exploitable security bugs), execution unstoppability, etc. We also make a standard assumption on blockchain security that the blockchain is immutable, fork-consistent, and Sybil-secure. The underlying security assumption is that a deployed blockchain system runs among a large number of peers with an honest majority, and compromising the majority of peers is hard.
This work is built on Ethereum's smart-contracts, cost model, and transaction model.
It treats Ethereum's consensus and underlying P2P networks as a blackbox.

\vspace{-0.1in}
\subsection{Design Goals \& Baselines}
\label{sec:design:etherbatch}

The design goal of \sysname{} is this: Through batching invocations, there should be a significant portion of the transaction cost saved (1. cost saving) for calling generic smart contracts (2. generalizability), while staying secure against the newly introduced adversary of off-chain \texttt{Batcher} (3. security). Specifically, the cost-saving goal is to reduce a significant portion of the Gas cost per invocation, via batching calls under the constraint of maximal transaction size. The generalizability goal is that the system should work with the general case of Batch Type S4. The security goal is to detect and prevent attacks mounted by the untrusted \texttt{Batcher} and protect the integrity of invocation information.
%YYY; we discuss the applicability of \sysname{} to other blockchain systems in \S~\ref{sec:YYYapplicability}.

There is limited research on batching smart-contract invocations. In Table~\ref{tab:rw}, we compare \sysname's research goal with other research work (i.e., Airdrop batching~\cite{DBLP:conf/esorics/FrowisB19}) and two baseline designs, which we will describe next. Here, ``no rewrite'' means no need to modify the smart contracts deployed for a DApp, for the ease of deployment.

\vspace{-0.05in}
\begin{table}[!htbp] %force in current page, disable float.
\caption{\sysname's design choices and related works}
\label{tab:rw}\centering{\small
\begin{tabularx}{0.375\textwidth}{ |X|c|c| }
  \hline
   & Generalizable & Cost Saving 
 \\ \hline
Baseline B1 & \xmark & \xmark 
 \\ \hline
Baseline B2 & \cmark & \xmark
 \\ \hline
\sysname{} & \cmark & \cmark
 \\ \hline
\end{tabularx}
}
\end{table}
\vspace{0.05in}

{\bf Baseline B1}: This baseline design of batching considers a special case. Suppose account $A$ is about to transfer tokens to $N$ other accounts $B_1,B_2,\dots,B_N$. Instead of sending $N$ transactions, account $A$ can set up a smart contract $C$ and send one transaction to $C$ that sends the $N$ transfers (e.g., by calling solidity's \texttt{transfer()} function $N$ times) in one shot. This is essentially the batching scheme used in existing works~\cite{DBLP:conf/esorics/FrowisB19} for airdropping tokens (a common practice to give away free tokens~\cite{me:airdrop:iosiro}). While this scheme handle the case of a single sender $A$, it can be naturally extended to support multiple senders $A_1,A2\dots$. In this case, multiple senders calls \texttt{approve} to delegate their account balance to a smart contract $C$ before $C$ can batch-transfer tokens to multiple receivers.

Overall, this batching scheme is limited as it depends on ERC20 functions (\texttt{approve}/\texttt{transfer}). Also it does not necessarily lead to cost saving, as each transfer still incurs at least one transaction (i.e., \texttt{approve}).

\ignore{
{\bf Baseline B2}: In this baseline design, named B2, we aim at supporting invocations to generic smart contracts. The \texttt{Batcher} batches a number of invocation requests to a batch translation and the \texttt{Dispatcher} smart contract extracts the invocations before relaying them to the callee smart contract. To defend against the invocation-manipulation attacks by the \texttt{Batcher}, the \texttt{Dispatcher} smart contract verifies the signatures of the original callers and also maintains the counters per caller accounts to detect replayed invocations. 

We implement the design and show that the overhead of \texttt{Dispatcher}, esp. in updating the per-account counters, offsets the cost saved by batching. 
}
}

{
\vspace{-0.1in}
\section{The \sysname{} Security Protocol}
\label{sec:baseline}
This section presents the design rational, protocol description, its security analysis, and the resultant system design. 
The full protocol analysis is described in \S~\ref{sec:securityanalysis}.

\vspace{-0.1in}
\subsection{Design Space: Security-Cost Tradeoff}

{\bf Batching framework}: We start by describing the design framework to support batching of invocations to generic smart contracts. In this framework, the \texttt{Batcher} batches a number of invocation requests and sends them in a batch translation to the \texttt{Dispatcher} smart contract. The \texttt{Dispatcher} extracts the invocations and relay them to the callee smart contracts. 

In our threat model, the \texttt{Batcher} mount invocation-manipulation attacks. To prevent a forged invocation, the \texttt{Dispatcher} verifies the signatures of the original callers. 

%To detect invocation omission, the untrusted \texttt{Batcher} is required to send the hash of batch transaction to callers who audit the inclusion of batch transaction in Ethereum blockchain and verify the presence of their invocations in the transaction. Note that invocation omission or generally denial of \texttt{Batcher} service is detected after the batch transaction is sent, and is not prevented before the batch transaction. To discourage the \texttt{Batcher} from omitting invocations, one can design an external incentive scheme in which a victim caller can present an evidence of misbehaving \texttt{Batcher} and receive penalty. 
%This incentive scheme is described in YYY.

\label{sec:baseline:b2}
{\bf Baseline B2}: To prevent replaying an invocation, a baseline design (B2) is to elevate a blockchain's native replay protection into the smart-contract level. Specifically, existing blockchain systems defend against transaction replaying attacks by maintaining certain states on blockchain and check any incoming transaction against such states to detect replay. For instance, Ethereum maintains a monotonic counter per account, called nonce, and checks if the nonce in any incoming transaction increments the nonce state on-chain; a false condition implies replayed transaction. Bitcoin maintain the states of UTXO to detect replayed transactions. 

In B2, we implement per-account nonces in the \texttt{Dispatcher} smart contract and use them to check against incoming invocations, in order to detect replayed invocations. 

{\bf A cost observation}: In our preliminary cost evaluation on Ethereum, we found {\it a sweet spot that the batching framework without replay protection can lead to positive cost saving, while adding the baseline design (B2) of replay protection end up with a negative cost saving}. That is, the overhead of maintaining nonces in smart contracts in B2 offsets the cost saving by batching invocations.

Thus, in \sysname, we avoid placing nonces in \texttt{Dispatcher} and focus on an off-chain defense against invocation replaying. With an untrusted \texttt{Batcher}, we assume every caller is online for an extended period that covers the batch time window its invocation is submitted. We propose an off-chain protocol in which callers interactively sign a batch transaction. Note that there is an alternative design that callers audit batch transactions after they are acknowledged from the blockchain; however, the audit scheme does not prevent (only detects) a replayed invocation. 

%YYY cost evaluation

\vspace{-0.1in}
\subsection{Protocol Description}
\label{sec:protocol}

The protocol supports the general-case batching, that is, batching Type S4 invocations.
Suppose in a batch time window, there are $N$ invocations submitted from different callers. The \sysname{} protocol follows the batching framework described above and it works in the following four steps:

1) In the batch time window, a caller submits the $i$-th invocation request, denoted by \texttt{call}$_i$, to the \texttt{Batcher} service. As in Equation~\ref{eqn:1:invocation}, the request \texttt{call}$_i$ contains the caller's address/public key \texttt{account$_i$}, callee smart contract address \texttt{cntr$_i$}, function name \texttt{func$_i$}, and argument list \texttt{args$_i$}. With $i\in[1,N]$, there are $N$ such invocations in the time window.

{
%\color{black}~\tangSide{Response to Review Q1b [\#258B/A/C]}
2) By the end of the batch time window, the \texttt{Batcher} prepares a batch message \texttt{bmsg} and sends to the callers for validation and signing. As shown in Equation~\ref{eqn:3:bmsg}, message \texttt{bmsg} is a concatenation of the $N$ requests, $\texttt{call}_i$'s, their caller nonces $\texttt{nonce}_i$'s, and {\it \texttt{Batcher} account's nonce, $\texttt{nonce}_{\texttt{B}}$}. Then, the \texttt{Batcher} broadcasts the batch message \texttt{bmsg} in parallel to all $N$ callers of this batch. Each of the callers checks if there is one and only one copy of its invocation \texttt{call}$_1$ in the batch message; specifically, this is done by checking equality between $\texttt{nonce}_1$ in the batch message and the nonce maintained locally by the caller. After a successful check of equality, the caller signs the message \texttt{bmsg\_sign}, that is, \texttt{bmsg} without callers' nonces as shown in Equation~\ref{eqn:4:bmsgsign}. The caller signs \texttt{bmsg\_sign} using the private key in her Ethereum account. She then send her signature to the \texttt{Batcher}. This step finishes until all $N$ callers have signed the message and return their signatures to the \texttt{Batcher}.
}

3) \texttt{Batcher} includes the signed batch message in a transaction's data field and sends the transaction, called batch transaction, to be received by the \texttt{Dispatcher} smart contract.This is presented in Equation~\ref{eqn:2:batch} where \texttt{CA} is the address of smart contract \texttt{Dispatcher}.

{\small
\begin{eqnarray}
\forall{i}, \texttt{call}_i &=&\langle{} \texttt{account}_i, \texttt{cntr}_i, \texttt{func}_i, \texttt{args}_i\rangle{}
\label{eqn:batchtx}
\label{eqn:1:invocation}
\\ \nonumber
 \texttt{bmsg} &=& \texttt{call}_1\| \texttt{nonce}_1 \| \texttt{call}_2 \| \texttt{nonce}_2 \|\dots \\ 
&& \|\texttt{call}_N\| \texttt{nonce}_N \| \texttt{nonce}_{\texttt{B}} \label{eqn:3:bmsg} \\ 
 \texttt{bmsg\_sign} &=& \texttt{call}_1\| \texttt{call}_2 \|\dots\| \texttt{call}_N \| \texttt{nonce}_{\texttt{B}} \label{eqn:4:bmsgsign} \\ \nonumber
 \forall{i}, \texttt{sig}_i &=&  sign_{\texttt{account}_i}(\texttt{bmsg\_sign}) \\ \nonumber
 \texttt{bsig} &=&  \texttt{sig}_1\| \texttt{sig}_2 \| \dots\|  \texttt{sig}_N \\ \nonumber
 \texttt{data} &=& \langle{}\texttt{dispatch\_func}, \texttt{bmsg\_sign} \| \texttt{bsig}\rangle{}  \\ \nonumber
 \texttt{tx} &=& \langle{} \texttt{account}_{\texttt{B}}, \texttt{nonce}_{\texttt{B}}, \texttt{CA}_{\texttt{D}}, \\
&& \texttt{sig}_{\texttt{B}}, \texttt{value}, \texttt{data}\rangle{}
\label{eqn:2:batch}
\end{eqnarray}
}

\begin{figure}[!bhtp]
\centering
\includegraphics[width=0.45\textwidth]{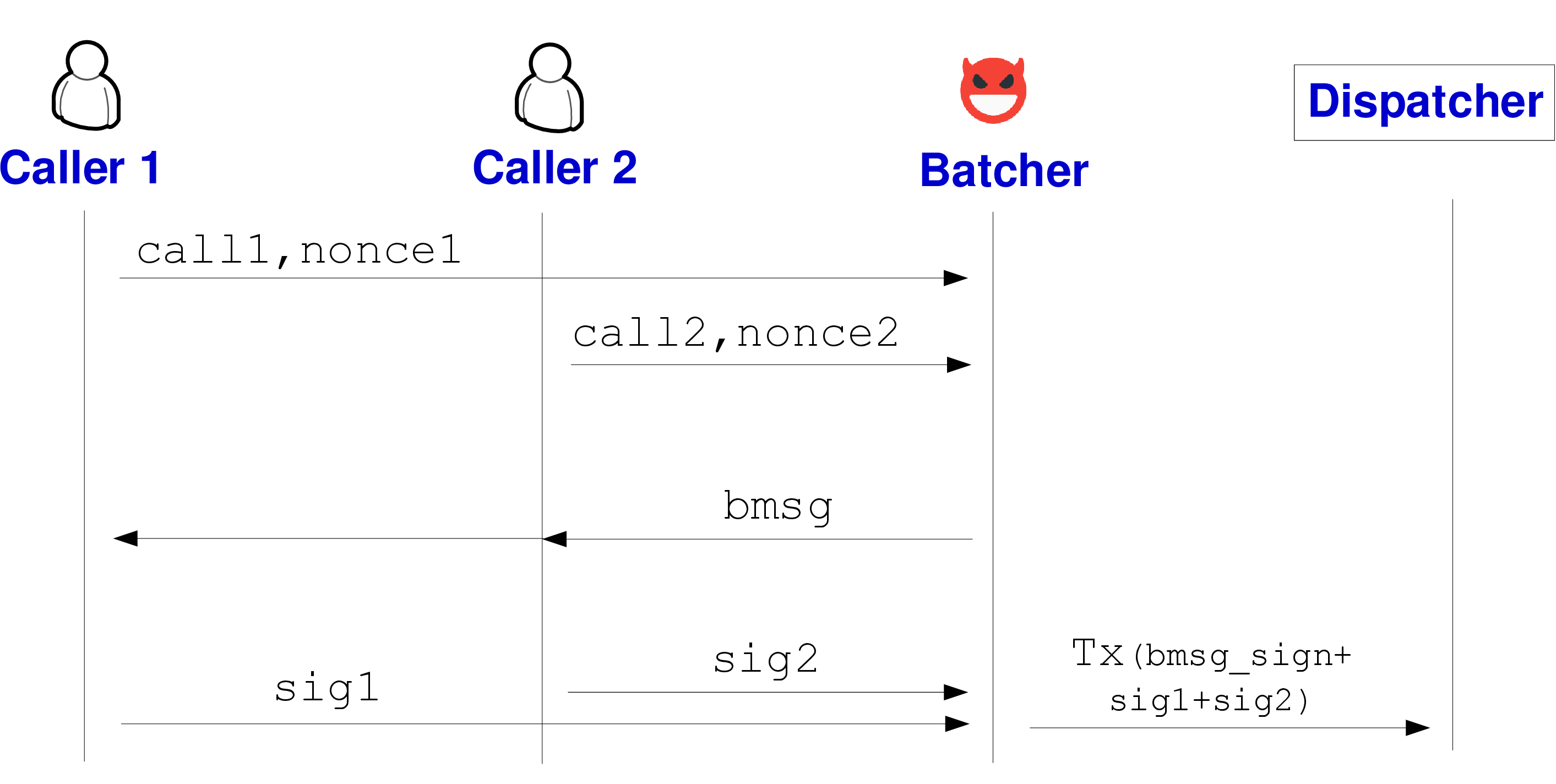}
  \caption{Generation of batch transaction off-chain among \texttt{Batcher} and two caller accounts}
  \label{fig:example:offchainsign}
\vspace{-0.1in}
\end{figure}

On the blockchain, 4) in function \texttt{dispatch\_func}, smart contract \texttt{Dispatcher} parses the transaction and extract the original invocations \texttt{call$_i$} before forwarding them to callees, namely \texttt{cntr$_i$} and \texttt{func$_i$}.
%XXX The pseudo code of \texttt{Dispatcher} is in Appendix~\ref{sec:code:dispatcher}. 
Particularly, smart contract \texttt{Dispatcher} internally verifies the signature of each extracted invocation against its caller's public key; this can be done by using Solidity function $\texttt{ecrecover}(\texttt{call}_i, \texttt{sig}_i, \texttt{account}_i)$. If successful, the \texttt{Dispatcher} then internal-calls the callee smart contract. At last, the callee function executes the body of the function under the given arguments \texttt{args}$_i$.
{
\color{black}
The pseudo-code of the smart contract \texttt{Dispatcher} is shown in Listing~\ref{lst:dispatch}
}

{\bf An example}: We use an example to illustrate the interactive signing process. In the example, there are $N=2$ callers respectively sending two invocations. The process causes five messages among the two callers and the \texttt{Batcher} off-chain and is illustrated in Figure~\ref{fig:example:offchainsign}.
{
\color{black}
\label{lst:dispatch}
\definecolor{mygreen}{rgb}{0,0.6,0}
\lstset{ %
  backgroundcolor=\color{white},   % choose the background color; you must add \usepackage{color} or \usepackage{xcolor}
  %basicstyle=\footnotesize\ttfamily,        % the size of the fonts that are used for the code
  basicstyle=\scriptsize\ttfamily,        % the size of the fonts that are used for the code
  breakatwhitespace=false,         % sets if automatic breaks should only happen at whitespace
  breaklines=true,                 % sets automatic line breaking
  captionpos=b,                    % sets the caption-position to bottom
  commentstyle=\color{mygreen},    % comment style
  deletekeywords={...},            % if you want to delete keywords from the given language
  escapeinside={\%*}{*)},          % if you want to add LaTeX within your code
  extendedchars=true,              % lets you use non-ASCII characters; for 8-bits encodings only, does not work with UTF-8
  %frame=single,                    % adds a frame around the code
  keepspaces=true,                 % keeps spaces in text, useful for keeping indentation of code (possibly needs columns=flexible)
  keywordstyle=\color{blue},       % keyword style
  language=Java,                 % the language of the code
  morekeywords={*,...},            % if you want to add more keywords to the set
  numbers=left,                    % where to put the line-numbers; possible values are (none, left, right)
  numbersep=5pt,                   % how far the line-numbers are from the code
  numberstyle=\scriptsize\color{black}, % the style that is used for the line-numbers
  rulecolor=\color{black},         % if not set, the frame-color may be changed on line-breaks within not-black text (e.g. comments (green here))
  showspaces=false,                % show spaces everywhere adding particular underscores; it overrides 'showstringspaces'
  showstringspaces=false,          % underline spaces within strings only
  showtabs=false,                  % show tabs within strings adding particular underscores
  stepnumber=1,                    % the step between two line-numbers. If it's 1, each line will be numbered
  stringstyle=\color{mymauve},     % string literal style
  tabsize=2,                       % sets default tabsize to 2 spaces
%  title=\lstname,                  % show the filename of files included with \lstinputlisting; also try caption instead of title
  caption = {Implement \texttt{Dispatcher} in smart contract},
  label = {lst:dispatch},
  moredelim=[is][\bf]{*}{*},
}
\begin{lstlisting}
contract Dispatcher {
function dispatch(uint256[] contractAddrs,uint256[] funcHashs,uint256[][] args,bytes[] sigs){
  for(int i=0; i<contractAddrs.length; i++){
    if(args[i].length==1){
      byte32 msgHash=keccak256(abi.encodePacked(contractAddrs[i],funcHashs[i],*args[i][0]*));
      require(sigs[i].length=65);
      r=mload(add(sigs[i],32));
      s=mload(add(sigs[i],64));
      v=byte(0,mload(add(sigs[i],96)));
      uint256 origSender=ecrecover(msgHash,r,s,v);
      if(!origSender) continue;
      contractAddrs[i].call(funcHashs[i],origSender,*args[i][0]*);}
    if(args[i].length==2){
      byte32 msgHash=keccak256(abi.encodePacked(contractAddrs[i],funcHashs[i],*args[i][0],args[i][1]*)); //differ from Line 5 in one more argument
      ...//repeat the code from Line 6-11
      contractAddrs[i].call(funcHashs[i],origSender,*args[i][0],args[i][1]*);} //differ from Line 12 in one more argument
    //Other cases with longer argument lists (args[i].length>=3) can be similarly supported.
\end{lstlisting}

}
{
\subsection{Security Protocol Analysis}
\label{sec:securityanalysis}

%~\tangSide{Response to Review Q1 [\#258B/A/C]}
{\bf Security against invocation-forging \texttt{Batcher}}: 
Invocation forging refers to that given a caller $A$ who did not send an invocation $X$, the \texttt{Batcher} forges the invocation $X$ and falsely claims it is sent by caller $A$. 
In \sysname, the hardness of \texttt{Batcher} making \texttt{Dispatcher} accept a forged invocation can be reduced to the hardness of forging a digital signature (as in Protocol Step 3) in \S~\ref{sec:protocol}), which is known to be with negligible probability. 

{\bf Security against invocation-omitting \texttt{Batcher}}: 
Invocation omission refers to that the \texttt{Batcher} omits an invocation in a batch while falsely acknowledging the victim client the inclusion of her invocation. In \sysname, an omitted invocation in a batch transaction included in the blockchain cannot be concealed from the victim client.
To prove it, omitting an invocation and concealing it from the client requires producing a sufficient number of fake blocks (e.g., 6 blocks in Bitcoin) where one of the blocks includes a fake transaction that includes the omitted invocation. Thus, this is equivalent to mounting a successful double-spending attack on the underlying blockchain, which is assumed to be hard. 

In addition to detectability, \sysname can be extended with an external incentive scheme (similar to IKP~\cite{DBLP:conf/sp/MatsumotoR17}) to punish a misbehaving \texttt{Batcher} and prevent her future omission of invocations.

%If the \texttt{Batcher} omits an invocation in the batch transaction but does not omit it in the batch message (\texttt{bmsg}), the \texttt{Dispatcher} will not be able to verify callers' signatures as the omitted invocation is included in the signatures (\texttt{bsig}). If the \texttt{Batcher} omits an invocation in the batch message, the victim caller will detect it and will not sign the batch message.

{\bf Security against invocation-replaying \texttt{Batcher}}: 
Invocation replaying refers to that the \texttt{Batcher} replays an invocation in a successful batch transaction without informing the victim client. There are different forms of replaying attacks, including R1) the \texttt{Batcher} replaying invocations twice (or multiple times) in the same batch transaction, R2) the \texttt{Batcher} replaying a batch transaction with the same nonce $\texttt{nonce}_{\texttt{B}}$, R3) the \texttt{Batcher} replaying a batch transaction's data twice with two different $\texttt{nonce}_{\texttt{B}}$, and R4) the \texttt{Batcher} intentionally generating smaller batches. Here, we don't consider the case of the \texttt{Batcher} replaying an invocation in two different batch transactions, in which one replayed copy must be an forged invocation to the caller and which can thus be prevented.

Overall, \sysname prevents invocation replaying in forms of R1, R2, R3 and R4. The following is the security analysis.

Consider R1 that a replayed invocation cannot appear in the first round message (\texttt{bmsg}), as the victim client can easily detect it and refuse to sign the joint message in the second round. If an invocation is replayed in the batch transaction, the \texttt{Batcher} has to modify the jointly signed message (\texttt{bmsg\_sign}) and forge all the second-round signatures, known to be hard. 

%If the \texttt{Batcher} replays an invocation in the batch transaction, the batch transaction will be different from the batch-signed message (which includes one and only one copy of invocation from a caller). And the difference will make Dispatcher's verification fail.

Consider Case R2 that the \texttt{Batcher} replays an entire batch transaction, that is, sending the batch transaction with the same nonce twice. Such a transaction-level replay will be prevented by Ethereum's native replay protection based on $\texttt{nonce}_{\texttt{B}}$. 

Consider Case R3 that the \texttt{Batcher} replays a batch transaction with different $\texttt{nonce}_{\texttt{B}}$. The \texttt{Dispatcher}'s verification will fail because the original $\texttt{nonce}_{\texttt{B}}$ is signed by callers (recall Equation~\ref{eqn:4:bmsgsign}).

Consider Case R4 that the \texttt{Batcher} may intentionally generate small batches; for instance, instead of one batch of $10$ invocations, it generates two smaller batches, each $5$ invocations. This is not necessarily an attack as the batch transaction size is bounded by Ethereum's native block Gas limit. But it could be a protocol deviation and can be detected: It will result in two batch transactions included in Ethereum at similar time (w.r.t., the batch time window). An auditing caller can detect the anomaly by inspecting the public Ethereum transaction and open disputes for further resolution.

{\bf Security against denial-of-service callers}: 
%~\tangSide{Response to Review Q1b [\#258B/A/C]}
\sysname can be extended to guarantee that a denial-of-service caller cannot delay the overall processing of a batch. 
In the extension, the \texttt{Batcher} enforces a timeout on waiting for callers' batch signatures. After the timeout, the \texttt{Batcher} generates the batch transactions, and \texttt{Dispatcher} does not forward to the callee smart contract an invocation whose batch signature is missing.
With this extension, a denial-of-service caller who delays her batch signature after the timeout will be ignored and does not invoke the callee smart-contract function, while other invocations are not affected. The DoS caller can only cause the fee of batch transaction to increase, which can be further detected and blacklisted by the \texttt{Batcher}.

This work does assume that the \texttt{Batcher} is always available. In practice, we consider this is a reasonable assumption as such a service can be run on highly-available cloud platforms, and real-world transaction relay services such as infura.io that require clients to trust its availability are already operational and widely adopted.
The \texttt{Batcher} service has incentives to protect its business and defend against external denial-of-service attacks.

{\bf Security against caller impersonator in collusion w. \texttt{Batcher}}: 
%~\tangSide{Response to Review Q1c [\#258B/A/C]}
Recall Figure~\ref{fig:example:offchainsign} that normally, the \texttt{Batcher} sends to Caller 2 the batch message \texttt{bmsg} that includes Caller 1's public key $PK_1$ and her invocation \texttt{call}$_1$. Caller 2 simply verifies \texttt{call}$_1$ against the provided $PK_1$ and, if it passes, signs \texttt{bmsg} before returning it to \texttt{Batcher}.
The malicious \texttt{Batcher} may include in \texttt{bmsg'} an impersonator's invocation, that is, \texttt{call}$_1'$ and her public key $PK_1'$. In this case, Caller 2 still verifies \texttt{call}$_1'$ in the \texttt{bmsg} against $PK_1'$, which passes and leads to Call 2's signature on \texttt{bmsg'}. Message \texttt{bmsg'} is returned to and signed by \texttt{Batcher}, is further verified successfully by \texttt{Dispatcher}, and gets \texttt{call}$_1'$ forwarded to the callee smart contract.
The callee will handle the internal call sent from $PK_1'$ and leave the actual sender (i.e., $PK_1$) unharmed.}

}

\vspace{-0.1in}
\subsection{System Overview}

\begin{figure}[!bhtp]
\centering
  \includegraphics[width=0.35\textwidth]{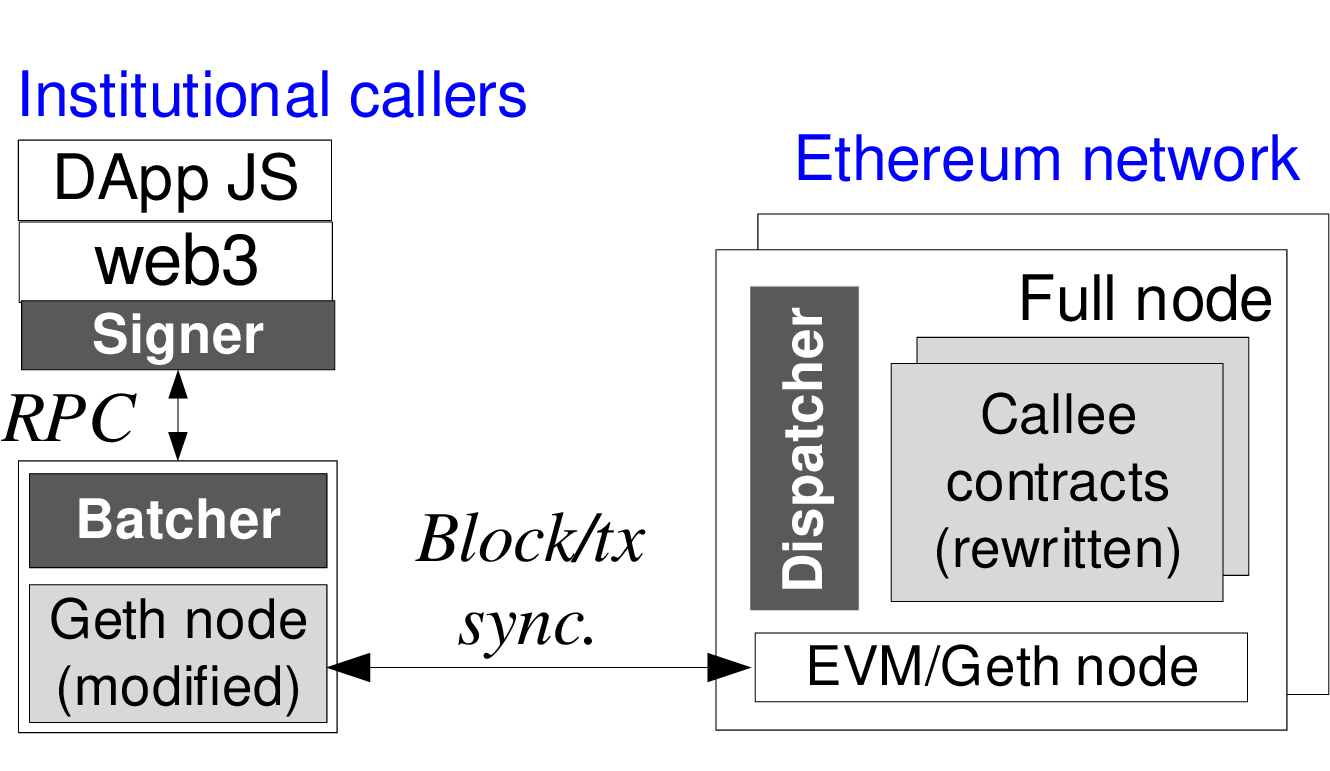}
  \caption{Retrofitting \sysname{} to Ethereum-based DApps: The right-hand side of this figure illustrates the general mechanism where the two dark shades are the core system components of \sysname, and the light shade is a statically instrumented Ethereum full node (running \texttt{Geth}).}
  \label{fig:model1}
\vspace{-0.1in}
\end{figure}

To materialize the protocol, we design a middleware system atop the underlying Ethereum-DApp ecosystem. Specifically, the system runs the \texttt{Batcher} middleware on an Ethereum node (e.g., a Geth client) that is synchronized with an Ethereum network. The \texttt{Dispatcher} smart contract runs on the Ethereum network and forwards invocations to the callee smart contracts. 
%XXX The pseudo code of the \texttt{Dispatcher} smart contract is presented in Listing~\ref{lst:dispatch} in \S~\ref{sec:code:dispatcher} in the Supplemental Material.

The off-chain \texttt{Batcher} is a middleware running on an untrusted third-party host. In general, the \texttt{Batcher} buffers incoming invocations submitted by callers and under certain conditions (as described below) triggers the batching of invocations. Once a batch of invocations is determined, the \texttt{Batcher} jointly works with original callers to generate the batch transaction (as described by the joint-signing process in \S~\ref{sec:protocol}).

\label{sec:hook}
{\bf Implementation}: To transparently support unmodified DApp clients, we statically instrument Geth's handling of raw transactions and expose hooks to call back the \texttt{Batcher}'s code that make decisions on batching, as will be described next. Specifically, the instrumented Geth node unmarshalls a raw transaction received, extracts its arguments, places it in \texttt{Batcher}' internal buffer (e.g., \texttt{bpool} as will be described) and makes essential decisions regarding which invocations to be included in the next batch transaction before actually generating and sending it (as described above). The statically instrumented Geth node retains the same \texttt{sendRawTransaction()}/\texttt{sendTransaction()} API and thus supports unmodified DApp clients. 
{
\color{black}
The pseudo-code showing how to hook \sysname{} into Geth is described in Listing~\ref{lst:instrument}
}

Next in \S~\ref{sec:customization}, we propose policies for \texttt{Batcher}'s decision-making that strikes balance between costs and delay. To integrate \sysname{} with legacy smart-contracts, we propose schemes to automatically rewrite smart contracts at scale, which is described in \S~\ref{sec:rewrite}.

{
\color{black}
\label{lst:instrument}
\lstset{ %
  backgroundcolor=\color{white},   % choose the background color; you must add \usepackage{color} or \usepackage{xcolor}
  %basicstyle=\footnotesize\ttfamily,        % the size of the fonts that are used for the code
  basicstyle=\scriptsize\ttfamily,        % the size of the fonts that are used for the code
  breakatwhitespace=false,         % sets if automatic breaks should only happen at whitespace
  breaklines=true,                 % sets automatic line breaking
  captionpos=b,                    % sets the caption-position to bottom
  commentstyle=\color{mygreen},    % comment style
  deletekeywords={...},            % if you want to delete keywords from the given language
  escapeinside={\%*}{*)},          % if you want to add LaTeX within your code
  extendedchars=true,              % lets you use non-ASCII characters; for 8-bits encodings only, does not work with UTF-8
  %frame=single,                    % adds a frame around the code
  keepspaces=true,                 % keeps spaces in text, useful for keeping indentation of code (possibly needs columns=flexible)
  keywordstyle=\color{blue},       % keyword style
  language=Java,                 % the language of the code
  morekeywords={*,...},            % if you want to add more keywords to the set
  numbers=left,                    % where to put the line-numbers; possible values are (none, left, right)
  numbersep=5pt,                   % how far the line-numbers are from the code
  numberstyle=\scriptsize\color{black}, % the style that is used for the line-numbers
  rulecolor=\color{black},         % if not set, the frame-color may be changed on line-breaks within not-black text (e.g. comments (green here))
  showspaces=false,                % show spaces everywhere adding particular underscores; it overrides 'showstringspaces'
  showstringspaces=false,          % underline spaces within strings only
  showtabs=false,                  % show tabs within strings adding particular underscores
  stepnumber=1,                    % the step between two line-numbers. If it's 1, each line will be numbered
  stringstyle=\color{mymauve},     % string literal style
  tabsize=2,                       % sets default tabsize to 2 spaces
  title=\lstname,                  % show the filename of files included with \lstinputlisting; also try caption instead of title
  caption={Hook \sysname to \texttt{Geth}},
  label={lst:instrument},
  moredelim=[is][\bf]{*}{*},
}
\begin{lstlisting}
//sendTx is the instrumented RPC 
//_sendTx is the original RPC
bool sendTransction(from,to,value,data){
  signedTx = sign(from,to,value,data);
  return sendRawTransction(signedTx);}
bool sendRawTranaction(signedTx){
  *from,to,value,sig,data = unmarshall(signedTx);*
  *Batcher.buffer(from,to,value,sig,data);*
  *if(Batcher.isFull()){*
    *batchtxs = Batcher.clearAllAndSerialize();*
    *_data = marshall(batchtxs);*
    *_from = Dispatcher_contract.owner;*
    *_to = Dispatcher_contract.address;*
    *_value = calValue(batchtxs);*
    return _sendTransction(_from,_to,_value,_data);}
  return true;}
\end{lstlisting}
}

{
\vspace{-0.1in}
\section{\texttt{Batcher}'s Policies}
\label{sec:customization}

In this section, we propose mechanisms and policies for the \texttt{Batcher} to properly batch invocations for design goals in cost and delay. We first formulate the design goal of optimizing Gas cost per invocation in the presence of the workload. We then formulate the design goal of reducing Ether cost per invocation without causing delay to when the invocation is executed on Ethereum.

\vspace{-0.1in}
\subsection{Optimizing Gas Cost}
\label{sec:opt:gas}

The degree of amortizing the cost by \sysname{} is dependent on the number and type of invocations put in a batch. In this subsection, we propose a series of policies that the \texttt{Batcher} can use in practice. The motivating observation is that there is no single policy that fits all (workloads), and under different workloads, the most cost-effective policy may differ. 

Note that the cost unit we consider here is Gas per invocation (which measures the amount of computational load an Ethereum node needs to carry out to serve an invocation). The proposed policies may cause invocation delay, and the policies are suitable for DApps that are insensitive to such delay.

\begin{itemize}
\item
{\bf $W$sec}: 
{\it Batching all invocations that arrive in a time window, say $W$ seconds}. 
In practice, the larger $W$ it is, the more invocations will end up in a batch and hence the lower Gas each invocation is amortized. However, a larger $W$ value means the \texttt{Batcher} needs to wait longer, potentially causing inconsistency and delay of invocation execution. We will systematically study the cost-delay tradeoff when taking into account the factor of Gas price in \S~\ref{sec:oneblock}.
\item {\bf Top$1$}: {\it Batching only the invocations that are sent from one account, such as the most intensive sender}. The motivation of doing this is that if all invocations needed batching are from one sender account, the batch transaction (of multiple invocations) only needs to be verified for once, thus eliminating the needs of verifying signatures in smart contracts and lowering the overhead. 

In practice, Top$1$ can be toggled on top of a $W$sec policy. For instance, $X$second-Top$1$ means batching only the invocations that arrive in a $W$-second window and are from the most intensive sender in that window. 

Whether the presence of Top$1$ batching policy can actually lead to positive Gas saving is dependent on workloads. If there is an institutional account sending invocations much more intensive than others, applying Top$1$ can lead to sufficient invocations in a batch and positive Gas saving. Otherwise, if the workload does not contain enough such invocations, the batch may be smaller than the one without Top$1$, which limits the degree of cost amortization. 

\item {\bf Min$X$}: {\it Only batch when there are more than $X$ candidate invocations in a batch time window.} The intuition here is that if there are too few invocations, the degree of cost amortization may be too low and can be offset by the batching overhead to result in negative cost saving. 
In 
%the full Technical Report~\cite{me:ibatch:tr},
\S~\ref{sec:costanalysis},
we conduct cost analysis based on Ethereum's Gas cost profile on different transaction operations and derive the minimal value of $X$ should be $5$. That is, it is only beneficial to generate a batch of at least $5$ invocations in a batch.
\end{itemize}

\vspace{-0.1in}
\subsection{Optimizing Ether Cost with Minimal Delay}
\label{sec:oneblock}

In this subsection, we consider a class of DApps, notably DeFi applications, that are sensitive to invocation timing. In these DApps, manipulating invocation timing or introducing invocation delay may cause consequences ranging from DApp service unresponsiveness to security damage (e.g., under the frontrunning attacks). Thus, we formulate the design goal to be optimizing Ether cost per invocation without introducing any invocation delay. We call the no-delay policy described in this subsection by 1block. Note that in Ethereum, the Ether cost of a transaction is the product of the transaction's Gas and its Gas price. 

Assume an oracle who can predict what invocations are included in a block (without batching) at the time when the invocations are submitted. An ideal, optimal offline algorithm is to batch the invocations in a future block and generate a batch transaction. If the Gas price of the batch transaction is set to be higher than at least one transaction in that future block, it is bound the batch transaction can be included in the same block with the unbatched case. In other words, no block delay is introduced. We call this approach by offline optimal batching as an ideal scheme.

In practice, the \texttt{Batcher} at the invocation submission time may not accurately predict when a block will be found and which block will include the invocation. We propose a realistic, online batching mechanism to reduce or eliminate the block delay.

{\bf Online batching w. minimal delay (1block)}: We propose a system design of \texttt{Batcher} atop an Ethereum client extending its memory pool (or \texttt{txpool}) functionality. We call this design by 1block.
We first describe the proposed system design and then decision-making heuristics. 
In a vanilla Ethereum client, a transaction is first buffered in memory (in a data structure called \texttt{txpool}), is then selected (by comparing its Gas price against other transactions in the \texttt{txpool}) by miners, and is included in the next block. 

In \sysname, the Ethereum client running on \texttt{Batcher} is extended with an additional memory buffer that we call \texttt{bpool} and that stores submitted invocations prior to the batch transaction. 

The \texttt{Batcher} service continuously receives the submitted invocations of registered DApps and buffer them into the \texttt{bpool}. To manage and evict invocations, the service periodically runs the following process: Every time it receives a block, the service waits for $d$ seconds and then executes Procedure \texttt{bpoolEvict} which produces a batch transaction to send to the Ethereum network. More specifically, the \texttt{bpoolEvict} procedure reads as input the transactions residing in the \texttt{txpool} and the invocations residing in the \texttt{bpool}. The procedure produces a batch transaction encoding selected invocations to be sent to the Ethereum network. There are two essential decisions to make by Procedure \texttt{bpoolEvict}: C1) What invocations to be evicted from \texttt{bpool} and to be put in the batch transaction. It also needs to decide C2) What Gas-price value should be set on the batch transaction. 

In addition to C1 and C2, the batching mechanism can be configured by $d$, that is, how long it waits after a received block to run Procedure \texttt{bpoolEvict}. In the following, we describe a series of policies to configure C1), C2) and $d$ of the \texttt{Batcher}.

{\bf Example}: We show an example process illustrated in Figure~\ref{fig:bpool}: It shows the timeline in which \texttt{bpool} on the \texttt{Batcher} operates and interacts with the remote Ethereum network. In the beginning ($0$-th second), the \texttt{Batcher} receives a block $B_0$ of 2 transactions, which evicts the 2 transactions from \texttt{txpool} and leaves it of $10$ transactions. Also assume there are $10$ invocations in the \texttt{bpool} in the beginning. On the $d=10$-th seconds, the service runs \texttt{bpoolEvict} which results in a batch transaction of $3$ invocations. It sends the batch transaction to the Ethereum network. As the Gas price of the batch transaction is high, it will be selected by the miners in the remote Ethereum network upon the next block $B_1$ being propagated, say on the $13$-th second. If the next-next block $B_2$ is found on the $20$-th second, the batch transaction will be included in $B_2$.

\begin{figure}[!bthp]
\centering
\includegraphics[width=0.425\textwidth]{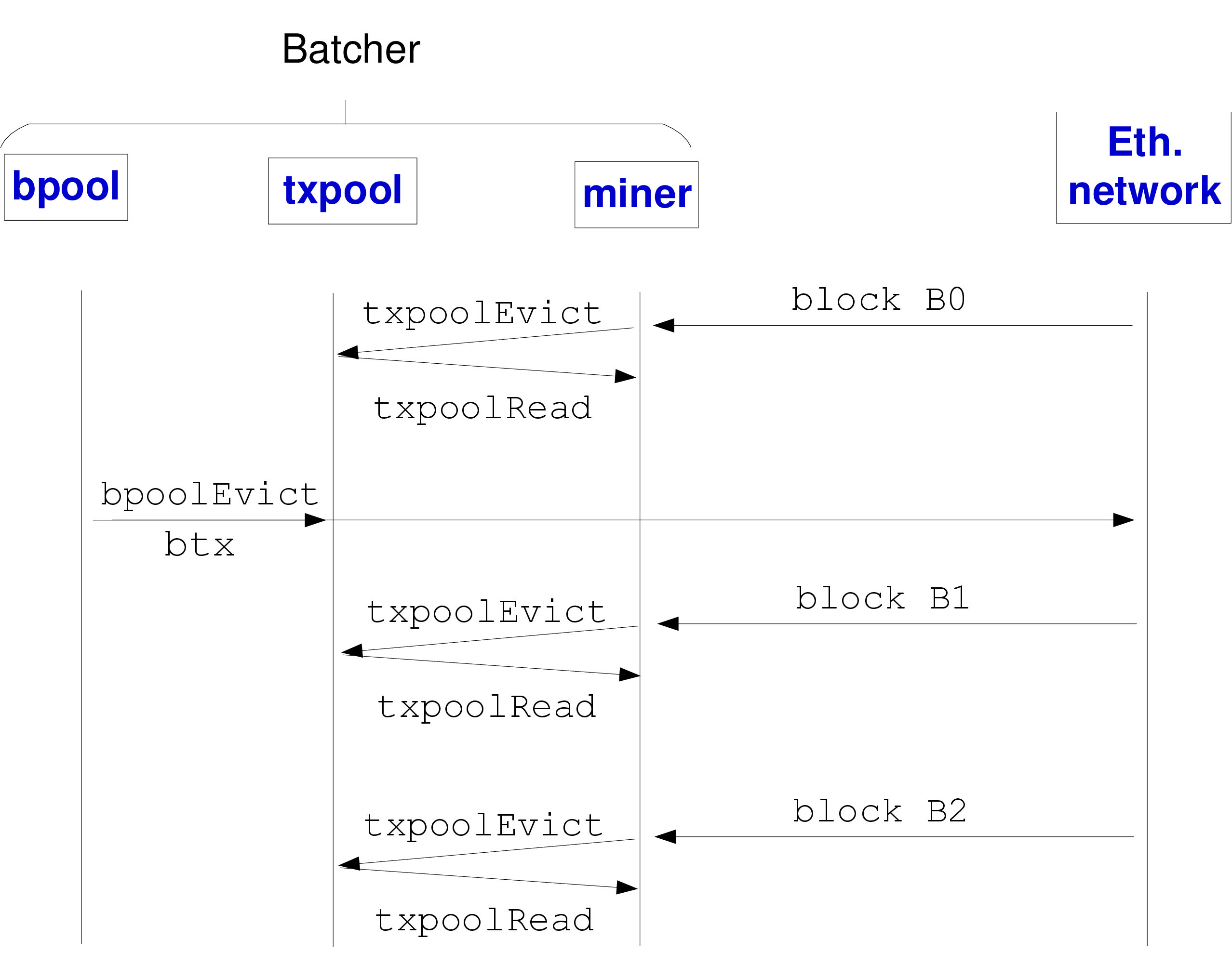}
\caption{An example process of running \texttt{bpool} and its eviction on \texttt{Batcher}.}
\label{fig:bpool}
\end{figure}

{\bf Heuristics}:
For C1), we propose to select the invocations in the \texttt{bpool} that have higher Gas price than $h$ such that the total Gas of transactions and invocations whose prices are higher than $h$ is under the block limit. Moreover, the total Gas of transactions and invocations whose prices are higher than $h-1$ is above the block limit.

{
%\color{black}
\label{sec:gaspricing}
For C2), a baseline is to set a fixed Gas price for every batch transaction, which does not reflect the price distribution in the current batch/block and can lead to excessive cost. We propose ``dynamic'' Gas pricing policies where the price of a batch transaction is dynamically set to ensure low Ether cost yet without delaying the block it will be included. We propose two policies:

\begin{itemize}
\item 
{\bf Batch-$X\%$}: 
The Gas price of a batch transaction is set to be above $X\%$ of the invocations in the batch.
\item
{\bf Block-$X\%$}: 
The Gas price of a batch transaction is set to be above $X\%$ of the transactions in the block (also including the invocations in the batch).
\end{itemize}

For instance, suppose there are 7 regular transactions included in a block and a batch transaction which consists of 3 invocations. The three invocations are associated with prices $8,9$ and ${10}$, and the 7 regular transactions' Gas prices are $1,2\dots{}7$. With policy batch-$50\%$, the batch transaction's price is $9$. With policy block-$10\%$, the batch transaction's price is 1. 
}

\ignore{
For C2), we use different approaches to set the batch transaction's Gas price. One can set the Gas price of the batch transaction in the same way with setting the Gas price of normal transaction. For instance, one can use the ``Fastest'' Gas price suggested by a Gas station.
Specific to the batch transaction, we propose to use the following as the batch transaction's Gas price: 1) The highest Gas price of the invocations in the batch, 2) the lowest Gas price of all transactions and invocations residing in \texttt{txpool} and \texttt{bpool} at the time of running \texttt{bpoolEvict}. 
}

\vspace{-0.1in}
\section{Integrating Legacy Smart Contracts via Rewriting}
\label{sec:rewrite}

When running legacy smart contracts on \sysname, the smart contracts need to be rewritten to authenticate the internal calls from \texttt{Dispatcher} smart contract. In this section, we
first describe two smart contracts rewriters that can automatically transform legacy smart contracts for supporting \sysname{} at scale. At the end, we talk about the integration of \sysname{} in future EVM probably without smart contracts rewriting.
\subsection{Source Code Rewriter}
\label{sec:rewrite:src}
The goal of our contract rewriting is to make an application smart contract accept the internal call by the \texttt{Dispatcher} contract. To do so, we design the following contract-rewriting procedure: Given an application smart contract \texttt{bar}, we create a new contract say \texttt{barByD} to inherent contract \texttt{bar}. We rewrite each function that contains references to \texttt{msg.sender}: Given such a function \texttt{foo(type original\_args)} in contract \texttt{bar}, we add in contract \texttt{barByD} a new function \texttt{fooByD(address from, type original\_args)}. 1) In this new function, a new argument \texttt{from} is added in function \texttt{fooByD}. The function body in \texttt{fooByD()} is the same with \texttt{foo()}, except for three modifications: 2) References \texttt{msg.sender} in \texttt{foo()} are replaced by argument \texttt{from} in \texttt{fooByD()}. 3) The first code line in  \texttt{fooByD()} asserts if the function caller is \texttt{Dispatcher}. 4) For any functions of \texttt{bar} that are called inside \texttt{foo}, the function invocation is rewritten to add a new argument \texttt{from}. In particular, this includes the case of modifier functions in solidity. Figure~\ref{lst:rewriting} illustrates the example of rewriting \texttt{transfer()} in an ERC20 token contract.

\definecolor{mygreen}{rgb}{0,0.6,0}
%\begin{scriptsize}
\lstset{ %
  backgroundcolor=\color{white},   % choose the background color; you must add \usepackage{color} or \usepackage{xcolor}
  %basicstyle=\footnotesize\ttfamily,        % the size of the fonts that are used for the code
  basicstyle=\scriptsize\ttfamily,        % the size of the fonts that are used for the code
  breakatwhitespace=false,         % sets if automatic breaks should only happen at whitespace
  breaklines=true,                 % sets automatic line breaking
  captionpos=b,                    % sets the caption-position to bottom
  commentstyle=\color{mygreen},    % comment style
  deletekeywords={...},            % if you want to delete keywords from the given language
  escapeinside={\%*}{*)},          % if you want to add LaTeX within your code
  extendedchars=true,              % lets you use non-ASCII characters; for 8-bits encodings only, does not work with UTF-8
  %frame=single,                    % adds a frame around the code
  keepspaces=true,                 % keeps spaces in text, useful for keeping indentation of code (possibly needs columns=flexible)
  keywordstyle=\color{blue},       % keyword style
  language=Java,                 % the language of the code
  morekeywords={*,...},            % if you want to add more keywords to the set
  numbers=left,                    % where to put the line-numbers; possible values are (none, left, right)
  numbersep=5pt,                   % how far the line-numbers are from the code
  numberstyle=\scriptsize\color{black}, % the style that is used for the line-numbers
  rulecolor=\color{black},         % if not set, the frame-color may be changed on line-breaks within not-black text (e.g. comments (green here))
  showspaces=false,                % show spaces everywhere adding particular underscores; it overrides 'showstringspaces'
  showstringspaces=false,          % underline spaces within strings only
  showtabs=false,                  % show tabs within strings adding particular underscores
  stepnumber=1,                    % the step between two line-numbers. If it's 1, each line will be numbered
  stringstyle=\color{mymauve},     % string literal style
  tabsize=2,                       % sets default tabsize to 2 spaces
  caption={Rewriting application contract. This figure uses the example of ERC20 token.}, % show the filename of files included with \lstinputlisting; also try caption instead of title
  label={lst:rewriting},
  moredelim=[is][\bf]{*}{*},
}
\begin{lstlisting}
//original functions
contract TokenOrig {
...
modifier noBlacklisted {
  assert(!isBlackListed[msg.sender]);_;}
function transfer(address to, unit256 value) noBlacklisted {
  super.transfer(to,value);
  balances[msg.sender] = SafeMath.safeSub(balances[msg.sender], value);
  balances[to] = SafeMath.safeAdd(balances[_to], value);}
//new functions added by iBatch
contract TokenByD is TokenOrig{
...
modifier noBlacklistedByD(address from) {
  assert(!isBlackListed[from]);}
function *transferByD*(*address from,*address to,unit256 value)
    *noBlacklistedByD(from)*{
  *assert(msg.sender!=dispatcher);*
  super.transferByD(from,to,value);
  balances[*from*]=SafeMath.safeSub(balances[*from*],value);
  balances[to]=SafeMath.safeAdd(balances[to],value);
}
\end{lstlisting}

\ifdefined\TTUT
//Inlined dispatcher contract
contract DispatcherInlined {
  ...
  mapping balances_12345_67890;
  mapping isBlackListed_12345_67890;
  function dispatch(uint256[] contractAddrs,uint256[] funcHashs,uint256[][] args,bytes[] sigs){
    ...
    uint256 origSender=ecrecover(msgHash,r,s,v);
    ...
    if(contractAddrs==12345 && funcHashs==67890){ //from.ERC20.transfer(to, value)
      require(!isBlackListed_12345_67890[origSender]);
      balances_12345_67890[origSender]=balances_12345_67890[origSender].sub(value);
      balances_12345_67890[to]=balances_12345_67890[to].add(value);
    } 
    ...
}
\fi

{
\color{black}
\subsection{Bytecode Rewriter}
\label{sec:rewrite:bytecode}
Because the majority of smart contracts deployed on Ethereum are without source code, we propose bytecode rewriting techniques. The goal is to facilitate the deployment of \sysname{} for these opaque smart contracts. Specifically, the bytecode rewriter will allow us to evaluate the Gas of \sysname{} on real opaque smart contracts, in a way to show the cost-effectiveness of \sysname{} to the owner of the contract for adoption.
Before we present our bytecode rewriter, we describe the preliminary of EVM bytecode layout.

{\bf Preliminary: EVM bytecode}: This work focuses on the representation of disassembled bytecode from remix~\cite{me:remix}. 

A smart contract is compiled into the bytecode format before being deployed into Ethereum via a transaction. In this contract-deploying transaction, the bytecode is layered out into creation code and runtime code. The transaction also includes the calldata storing arguments to invoke the contract constructor. The job of creation bytecode is to deploy runtime bytecode at an EVM address to be returned. To do so, the creation code accesses the constructor arguments to invoke the constructor. It also copies the runtime bytecode into EVM.

The data layout of an EVM smart contract includes a stack where data is directly accessed for instruction execution, a random-access memory, persistent storage as a key-value store, and invocation arguments (in calldata).
%By analogy to computer architectures, the EVM's stack/memory/storage is data in CPU/physical memory/storage medium.

The runtime bytecode handles incoming function calls (from transactions or other smart contracts) with a unified entry point at Instruction 0. Given function arguments in an invocation (or calldata), the call handling path 0) updates a pointer to freed memory space, checks the arguments, 1) runs function selector that maps hashed function signatures to the location storing the function wrapper, 2) executes the function wrapper which copies the function arguments from calldata into the stack, before 3) executing the function body.

{\bf Bytecode rewriting}: The goal is to implement \sysname's rewriting rules at the bytecode level. To do so, we systematically instrument both the runtime and creation bytecode.

{\it Rewriting runtime bytecode}: In the runtime bytecode, we modify the function-call handling path: 1) In the selector, we add a new conditional clause to forward the call on \texttt{transferByD} to its new wrapper. 
2) In the function wrapper, we add the code to push the new argument \texttt{from} in calldata to the top of the stack.
3) In the function body, because of the additional argument \texttt{from}, we modify the epilogue to destroy the arguments in the stack before returning properly. We replace msg.sender (or CALLER instruction) with the \texttt{from} argument from the stack. Normally, \texttt{from} can be referenced by the top of the stack. Some cases need special handling. For instance, when the msg.sender is used in the context of a function call, 
%such as Line 7 and 8 of Listing~\ref{lst:rewriting}, 
\texttt{from} may not be on the stack top. 
In this case, our instrumentation code locates the code block where msg.sender resides (recall that code block is a straight-line code that begins with \texttt{JUMPDEST} and ends with an instruction that changes the control flow). In the first line after \texttt{JUMPDEST}, it stores a copy of the current stack top, which is \texttt{from}, in the \texttt{memory} by leveraging the free memory pointer. Then, the msg.sender is replaced by \texttt{from}'s \texttt{memory} copy (instead of the stack copy).

If the function body calls into another function, say foo(), which references msg.sender, the function A may need to be instrumented. An example is that \texttt{transfer()} in an inherited contract calls its parent contract's \texttt{transfer()}.
%, as in Line 18 in Listing~\ref{lst:rewriting}. 
The parent contract's \texttt{transfer()} needs to be instrumented as well. For another example, \texttt{transfer()} may call \texttt{transferFrom()} in its body and in this case, there is no need to instrument \texttt{transferFrom()} which already contains argument \texttt{from}. To distinguish the two cases, a function that does not need rewriting would be the one that does not access msg.sender. Otherwise, we always add an argument \texttt{from} and use it to replace the occurrence of msg.sender. One exception is the case of \texttt{transferFrom()}; since we semantically know from the ERC20 standard that the first argument in \texttt{transferFrom} is \texttt{from} that can be reused to replace msg.sender in \texttt{transferFrom}, if any.

{\it Rewriting creation bytecode}: In addition to runtime bytecode, we rewrite the creation bytecode. Recall that the creation bytecode needs to copy the runtime code from the transaction to EVM and copy the function arguments from the transaction to the stack. Because the rewritten runtime code has changed in length, the two copy functions need instrumentation, and the source location of the copy needs to be adjusted.

%\color{black}~\tangSide{Response to Review Q7 [\#258C]}
This work rewrites only the directly called functions by \texttt{Dispatcher} (call depth 1) and the \texttt{delegatecall}'ed functions at call depth larger than 1. There is no need to rewrite the functions at a call depth larger than $1$ and that are not \texttt{delegatecall}'ed. 
%Note that the current \sysname implementation does not support smart contracts accessing \texttt{tx.origin}.

\subsection{Possible Integration to Future EVM}
\label{sec:eip3074}
%~\tangSide{Response to Q4' [\#258D.iii]}
We note that a recent Ethereum Improvement Proposal (i.e., EIP-3074~\cite{me:eip:3074}) may facilitate the integration of \sysname{} with legacy smart contracts. EIP-3074 adds new EVM instructions (\texttt{AUTH} and \texttt{AUTHCALL}) that allow a smart contract to send invocations on behalf of EOA accounts: If a so-called invoker smart contract \texttt{AUTHCALL}s a callee smart contract with a signature from an EOA $X$, the callee smart contract would treat the invocation as if its message sender is directly from $X$ (instead of the invoker smart contract). This EIP is currently in review. 

In the future, if EIP-3074 is adopted by the Ethereum protocol, it would allow integrating \sysname{} with legacy smart contracts without rewriting. That is, the \texttt{Dispatcher} smart contract can simply issue an \texttt{AUTHCALL} to the callee smart contract with the original caller's signature. An EVM with EIP-3074 would allows unmodified callee smart contracts to accept such \texttt{AUTHCALL} from \texttt{Dispatcher}.
It can greatly facilitate \sysname's adoption and integration with the large number of legacy smart contracts on the Ethereum mainnet. 
}

\vspace{-0.1in}
\section{Evaluation}
\label{sec:eval}
This section presents the evaluation of \sysname. We report \sysname's performance (cost and delay) in comparison with the unbatched baseline under real workloads. We formulate two research questions (RQ1 and RQ2) that are respectively answered by our experiments in \S~\ref{sec:eval:gas} and \S~\ref{sec:eval:etherdelay}. 
We present 
%Technical Report~\cite{me:ibatch:tr}
other experiments that answer research questions comparing \sysname{} with batched baselines in \S~\ref{appdx:sec:eval:micro}.
%:{\it RQ3: How much Gas per invocation does \sysname{} cause in comparison with all baselines including unbatched (B0) and batched baselines (B1 and B2), under synthetic workloads?}

\vspace{-0.1in}
\subsection{Evaluating Gas Cost}
\label{sec:eval:gas}

\begin{figure*}[!ht]
\begin{minipage}{.6\textwidth}
  \begin{center}
    \subfloat[batch size]{%
      \includegraphics[width=0.5\textwidth]{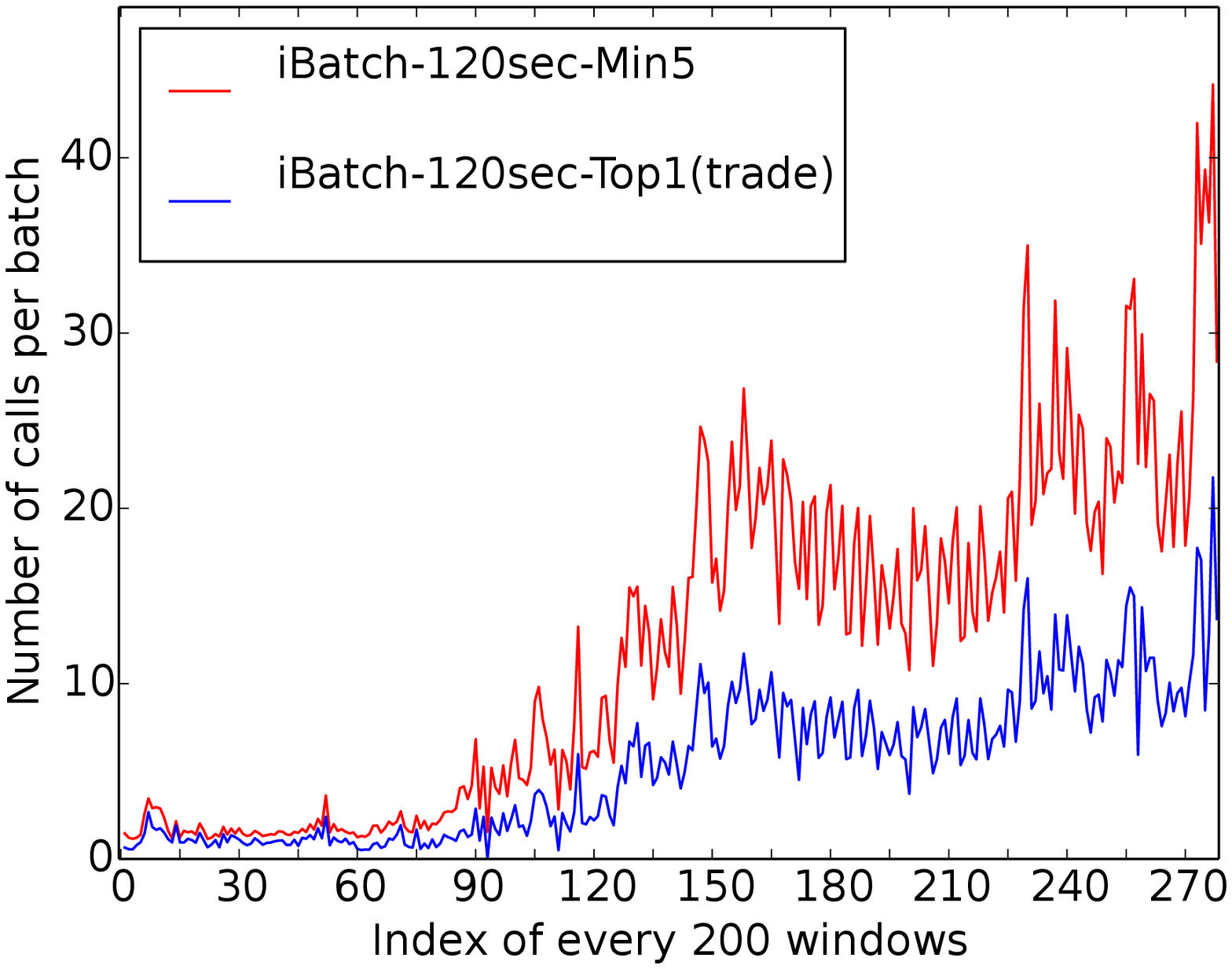}
      \label{fig:batchallsize}%
    }%
    \subfloat[gas cost per call]{%
      \includegraphics[width=0.5\textwidth]{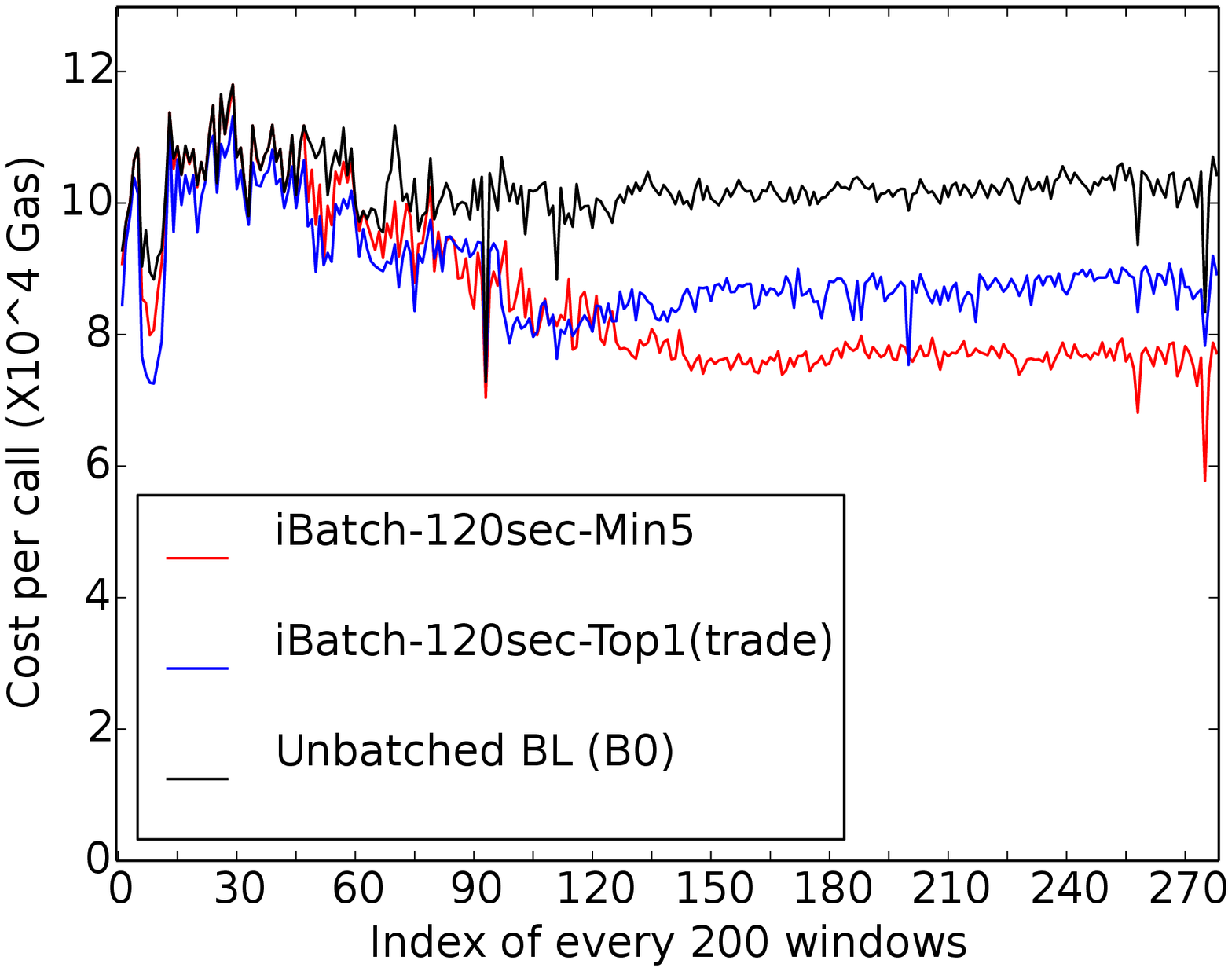}
      \label{fig:batchallcost}%
    }%
  \end{center}%\vspace{-0.15in}
  \caption{IDEX trace (5 months): 3 functions batch result}\label{fig:complex}\vspace{-0.10in}
\end{minipage}
\begin{minipage}{.39\textwidth}
%\begin{table}[!htbp] %force in current page, disable float.
\centering{\footnotesize
\begin{tabularx}{\textwidth}{ |l|X|l| }
  \hline
Traces & Policies & Gas per call (10k)  \\ \hline
\multirow{3}{*}{IDEX} 
%& \sysname-1block-min5 & $10.18$ ($-12.01\%$)\\ \cline{2-3}
& \sysname-120sec-min5 & $7.78$ ($-23.68\%$)\\ \cline{2-3}
& \sysname-120sec-top1 & $8.71$ ($-14.59\%$)\\ \cline{2-3}
& Unbatched BL (B0) & $10.20$\\ \hline
\multirow{3}{*}{BNB} 
&\sysname-120sec-min5 & $2.14$ ($-59.13\%$)\\  \cline{2-3}
&\sysname-120sec-top1 & $3.79$ ($-27.77\%$)\\  \cline{2-3}
&Unbatched BL (B0) & $5.25$\\ \hline
\multirow{2}{*}{Chainlink} 
%&\sysname-1block-min5 & $10.18$ ($-12.01\%$)\\  \cline{2-3}
&\sysname-120sec-min5 & $9.53$ ($-17.62\%$)\\  \cline{2-3}
&Unbatched BL (B0) & $11.57$\\ \hline
\end{tabularx} 
\caption{Average Gas cost per invocation}
\label{tab:avggas}
}
\end{minipage}
%\end{table}
\end{figure*}

{\it RQ1: How much Gas per invocation does \sysname{} result in, under different policies and in comparison with the unbatched baseline (B0), under real workloads?}

{\bf Motivation}: Gas per invocation is the metric directly affected by \sysname. This metric shows certain aspects of \sysname's cost-effectiveness. 
\sysname's Gas per invocation is sensitive to different policies (described in \S~\ref{sec:customization}). It is also dependent on the actual workload (e.g., how frequent invocations are sent in a fixed period). We set up this RQ to explore the sensitivity to policies and real workloads. 

{\bf Experiment methodology}: First, we choose three representative and popular DApps, that is, IDEX (representing decentralized exchange), BNB token (representing ERC20 tokens), and Chainlink (representing data feeds). We collect the DApps' invocations by running an instrumented Geth node to join the Ethereum mainnet. During the (basic) node synchronization, the node is instrumented to intercept all the transactions (i.e., external calls) and internal calls and dump them onto a local log file. 

Then, we prepare the collected trace to be replayable with accurate Gas cost. To do so, we replace the Ethereum addresses (i.e., public keys of account holders) by new public keys that we generated. This allows us to know the secret keys of the addresses used in the trace and use them to unlock the accounts (and sign transactions) during the replay. In addition, for cost-accurate replaying, we collect the pre- and post-states of relevant smart contracts of the DApps (e.g., BNB token balances) on Ethereum by crawling the website \url{https://oko.palkeo.com}.

In the experiments, we first unlock all senders' accounts, then replay the invocations with mining turned off, and at last turn on the miner to obtain the transaction receipt and Gas cost. This procedure does not require us to wait for transaction receipts, individually, and can greatly speed up the whole transaction-replaying process, especially in large-scale experiments. In this experiment, transactions/invocations in the original trace are replayed based on the block time, namely the block in which the transactions are originally included in real life. In the trace, only external calls are replayed and internal calls are used to cross-check the correctness of the replaying.  

{\bf Experiment settings}: 
We choose an IDEX trace that contains $664,863$ transactions calling three IDEX's functions: \texttt{deposit}, \texttt{trade} and \texttt{withdraw}.
The trace represents Ethereum transactions submitted from Sep. 2017 to Feb. 2018 (5-month long).
In the experiment, we replay the trace on our experiment platform, with and without \sysname. When running \sysname, we adopt two batching policies: 
1) Batch all invocations in each $120$-second window if there are more than $n_{min}=5$ invocations in that window. The policy is denoted by 120sec-min5.
2) Batch all \texttt{trade} invocations in each $120$-second window. The policy is denoted by 120sec-top1. 
Recall that given a time window, the top1 policy means batching only the invocations from the most popular caller in that window, which in this case is the IDEX2 or the caller of \texttt{trade}.
% Min$N$ policy means batching all invocations in the time window, only if there are more than $N$ invocations; when there are fewer than $N$ calls in a window, no batch transaction is generated.
Additionally, we set a maximal batch size to be $60$ invocations, so that the Gas of batched transaction does not exceed the block Gas limit.
In each experiment, we collect the resultant batch sizes and Gas cost of batched and unbatched transactions, from which we further calculate the Gas cost per call.

{\bf Results}: 
Figure~\ref{fig:batchallsize} shows the batch-size distribution over time. 
Each tick on the X axis is a time period of 200 windows (i.e., $200\cdot{}120$ seconds=$400$ minutes), and the Y value is the average size of the batches generated during that 200-window period. In the beginning, the generated batches are small, largely due to the fact that the distribution of calls are sparse. After the X index grows over 90, calls are more densely distributed and it generates larger batches. Comparing the two batching policies, the min5 policy generates batches that are $125\%$ larger than those generated by the top1 policy. This can be explained by that min5 policy considers all three functions in a batch and top1 considers only \texttt{trade} function, thus the former generates larger batches.

Figure~\ref{fig:batchallcost} illustrates the average Gas per call over time.
In the beginning, the two \sysname{} and the unbatched baseline B0 result in similar per-call costs, because of sparse call distribution over time and no chance of generating batches. After the X index grows over 90, it becomes clear that the \sysname{} under min5 results in the lowest Gas per call, which is $23.68\%$ smaller than that of unbatched baseline (B0). The \sysname{} under top1 results in a Gas per call that is $14.59\%$ lower than that of B0.

From these two figures, we summarize the average Gas per call in the first three rows of the table in Figure~\ref{tab:avggas}. We conducted similar experiments under the other DApps' trace and show the \sysname's performance in the rest of the table. 
{
%\color{black}~\tangSide{Response to Q6' [\#258D.i]}
Specifically, the BNB trace is from July 7, 2017 for 8 months, and the Chainlink trace is from Oct. 1, 2020 to Dec. 27, 2020.}
It can be seen that at the batch time window of 120 seconds, \sysname{} can generally save $14.59\sim{}59.13\%$ Gas cost per call compared with the unbatched baseline (B0).  

{
%\color{black} 
\vspace{-0.1in}
\subsection{Evaluating Ether Cost \& Delay}
\label{sec:eval:etherdelay}

\begin{figure}[!ht]
  \begin{center}
    \subfloat[Ether cost over time (each point is a $75$-second period).]{%
      \includegraphics[width=0.25\textwidth]{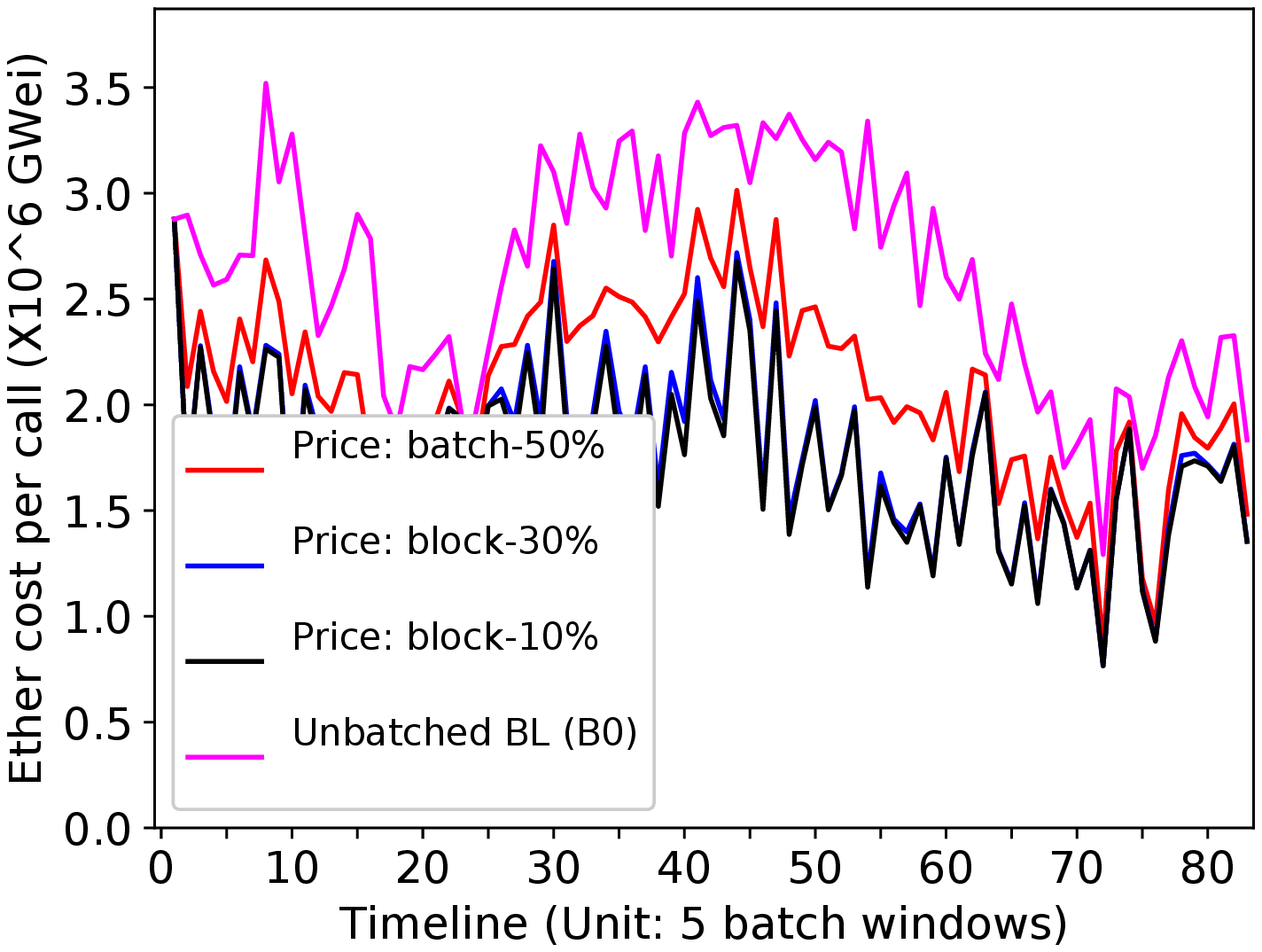}
      \label{fig:onlinecost15}%
    }%
    \subfloat[Block delay distribution (with $W=15$-second windows).]{%
      \includegraphics[width=0.25\textwidth]{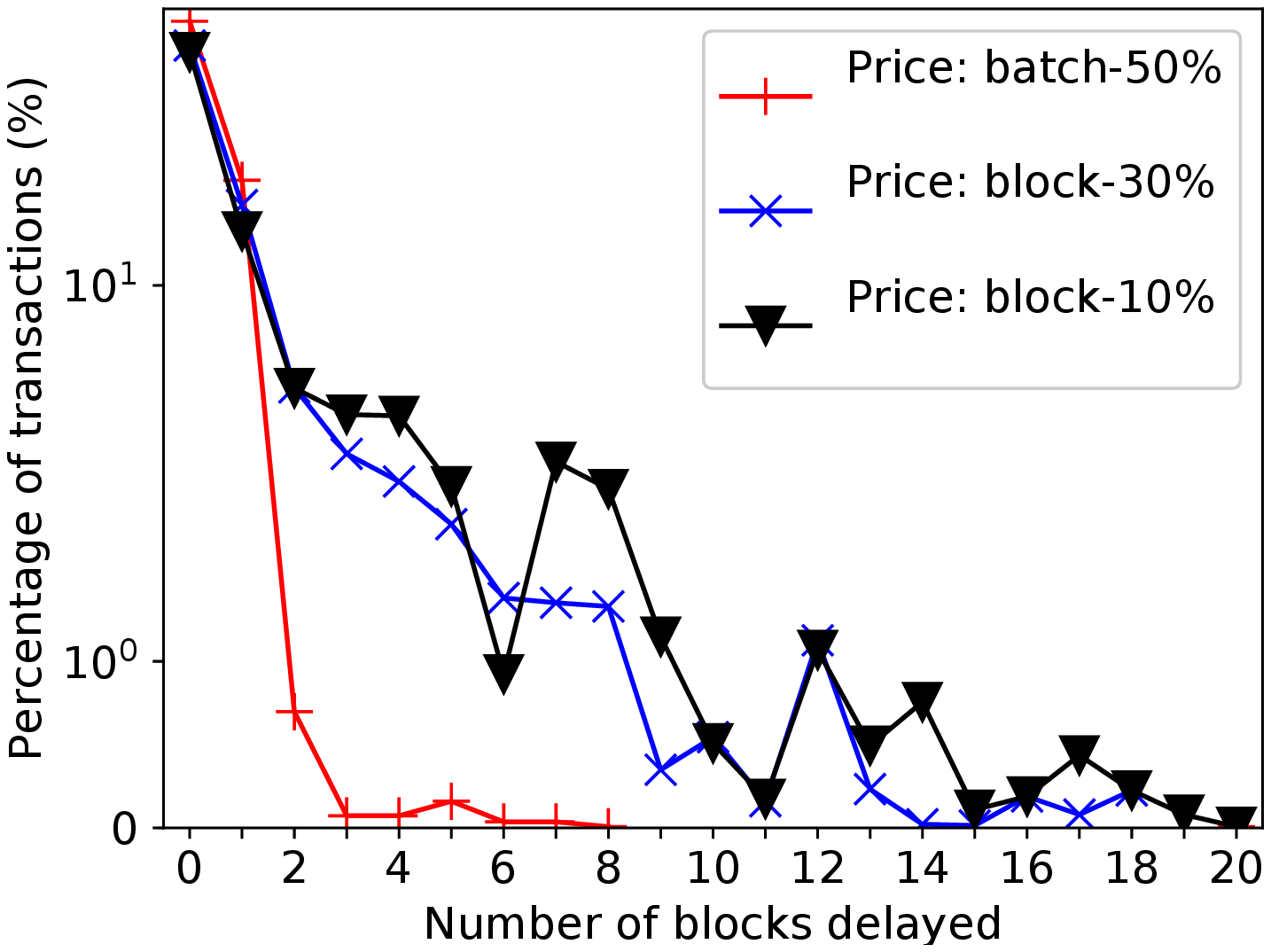}
      \label{fig:onlinedelay15}%
    }%
    \end{center}%\vspace{-0.15in}
  \caption{15 seconds window cost and delay}\label{fig:win15}\vspace{-0.10in}
\end{figure}

\begin{figure}[!ht]
  \begin{center}
      \includegraphics[width=0.275\textwidth]{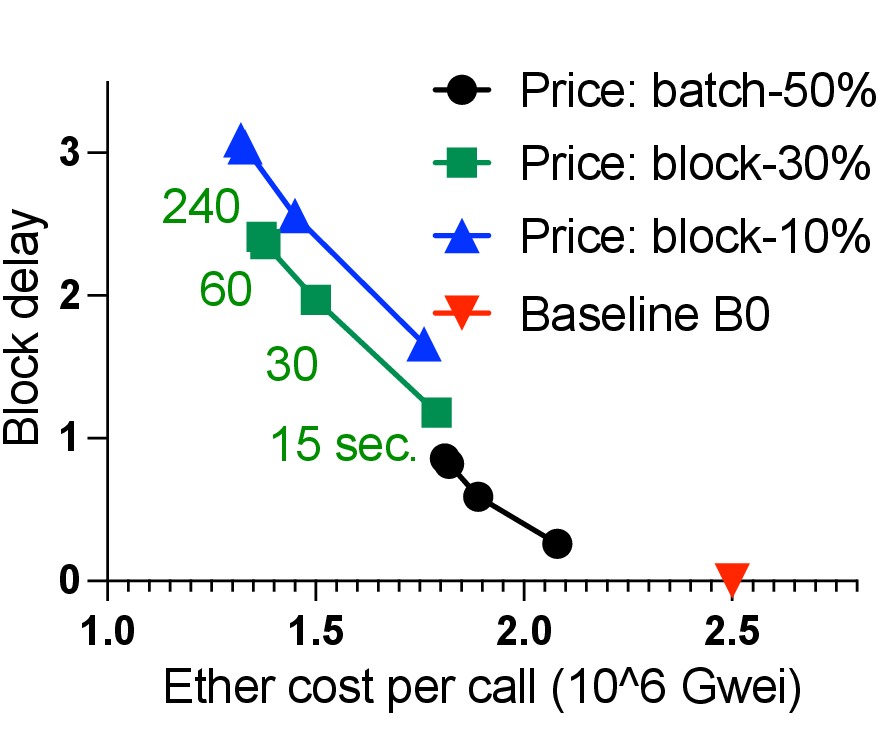}
    \end{center}%\vspace{-0.15in}
  \caption{Tradeoff between Ether cost and block delay under varying Gas price and batch window}
  \label{fig:costdelay}%
\end{figure}

\ignore{
\begin{minipage}{.45\textwidth}
\centering{\footnotesize
  \begin{tabularx}{\textwidth}{ |X|X|X|l| }
  \hline
  Settings ($W$-price) & $10^4$ Gas per call & $10^6$ GWei per call & Block delay \\ \hline
  15-10 & $3.50$ ($-13.79$\%)  & $1.76$ ($-31.52$\%) & $1.66$\\ \hline
  15-30 & $3.50$ ($-13.79$\%)  & $1.79$ ($-30.35$\%) & $1.18$\\ \hline
  15-Med & $3.50$ ($-13.79$\%) & $2.08$ ($-19.06$\%) & $0.26$\\ \hline
  30-Med & $3.29$ ($-18.97$\%) & $1.89$ ($-26.46$\%) & $0.59$\\ \hline
  60-Med & $3.18$ ($-21.65$\%) & $1.82$ ($-29.18$\%) & $0.82$\\ \hline
  240-Med & $3.17$ ($-21.92$\%) & $1.81$($-29.72$\%) & $0.86$\\ \hline
  B0 & $4.06$ & $2.57$ & $0$\\ \hline
  \end{tabularx}
  }
  \caption{Average Ether cost per call and delay under the online Tether trace}
  \label{tab:onlinetether}
\end{minipage}
}

{\it RQ2: How to characterize the Ether-delay tradeoff attained by different batching policies?
And how much Ether cost per invocation can \sysname save while with minimal block delay (compared with unbatched baseline B0)?}

{\bf Motivation}: On Ethereum, the cost metric that an end user (Ether owner) cares the most is the amount of Ether she needs to pay out of pocket for invocations. The Ether cost per invocation is the product of the Gas of an invocation and the Gas price of the (batch) transaction. RQ2 focuses on measuring the Ether cost per invocation.

Many DeFi applications are very sensitive to the timing of invocations, that is, when an invocation is included in the blockchain. Additional delay to the invocation may invite loss of financial opportunity (e.g., in an auction), increase exploitability under frontrunning attacks, et al. We mainly use the 1block online mechanism (in \S~\ref{sec:oneblock}) that causes minimal block delay to batched invocations.

{\bf Experiment methodology}: We follow the same transaction-replaying method described before, with the only exception: To measure delays under 1block, we have to know each transaction's submission time. This is obtained by crawling the transaction's ``pending'' time from website \url{etherscan.io}  (an example link is ~\cite{me:etherscan:tx}). Then, a transaction's submission time is its block time minus the pending time.

{\bf Experiment settings}: We collect a trace of $100,000$ Ethereum transactions, each invoking Tether's \texttt{transfer()} function~\cite{me:tether}. In real life, these transactions were submitted in one day on Oct. 4, 2020. We did not collect more transactions as replaying $100,000$ transactions takes around $570$ minutes, which is long enough for conducting our experiments.

We replay the transaction trace in the following manner: We apply a pre-configured batching policy to generate a batch transaction, say at time $t$. How a block is produced and which transactions will be included in a block are simulated in the following manner (an approach also used in~\cite{DBLP:journals/corr/abs-2101-08778}): Given a specified Gas price $p$, the batch transaction submitted at time $t$ will be included in the first block produced after $t$ which includes at least one transaction with Gas price lower than $p$.

Following the above method, we replay the trace with \sysname with 1block mechanism and under different batching policies.

{\bf Results}: When replaying the trace, we use three pricing policies, namely batch-$50\%$, block-$30\%$ and block-$10\%$, as described in \S~\ref{sec:gaspricing}.
We measure each transaction's Gas and multiply it with its Gas price to obtain the transaction's Ether cost. By summing the Ether costs of the transactions in a unit time period and dividing it with the number of calls, we report the average Ether cost per call in Figure~\ref{fig:onlinecost15} where the unit period is $5$ windows (or $5\times{}15=75$ seconds). The results shows that \sysname of policy block-$10\%$ achieves the lowest Ether cost, which is 31.52\% lower than that of the unbatched baseline (B0). By comparison, \sysname under the batch-$50\%$ policy saves $19.06\%$ Ether per invocation than the baseline B0. 

We also plot the block delays of \sysname under these three configurations in Figure~\ref{fig:onlinedelay15}. The figure shows the distribution of batch transactions in their block delays. As can be seen, under the batch-$50\%$ policy, majority of the batch transactions have a minimal delay under three blocks. In average, the delay of \sysname under the pricing of $batch-50\%$ is 0.26 blocks, the delay under the price of 30 Gwei per Gas is 1.18 blocks, and the delay under the price of 10 Gwei per Gas is 1.66 blocks.

We then report the tradeoff between block delay and Ether cost per call under varying Gas prices of batch transaction and batch windows. The result is in Figure~\ref{fig:costdelay}. It can be seen with the batch transaction of the same Gas price (i.e., block-$30\%$ in the figure), the block delay increases and Ether per call decreases as the batch window grows from $15$ seconds through $240$ seconds. The unbatched baseline B0 incurs $0$ block delay and $2.5\times{}10^6$ Gwei per call. In comparison to the baseline, with the batch-$50\%$ policy and 15-second batch window, \sysname saves $19.06\%$ cost at the expense of delaying invocations by 0.26 blocks. With the policy of block-$10\%$ and $15$-second batch window, \sysname saves $31.52\%$ cost at an average $1.66$ block delay.
}
{
\color{black} 
\subsection{Gas Evaluation under Synthetic Workloads}
\label{appdx:sec:eval:micro}

{\it RQ3: How much Gas per invocation does \sysname cause in comparison with all baselines including unbatched (B0) and batched baselines (B1 and B2), under synthetic workloads?}

\begin{figure}
\centering
  \subfloat[Comparison to baselines]{%
    \includegraphics[width=0.25\textwidth]{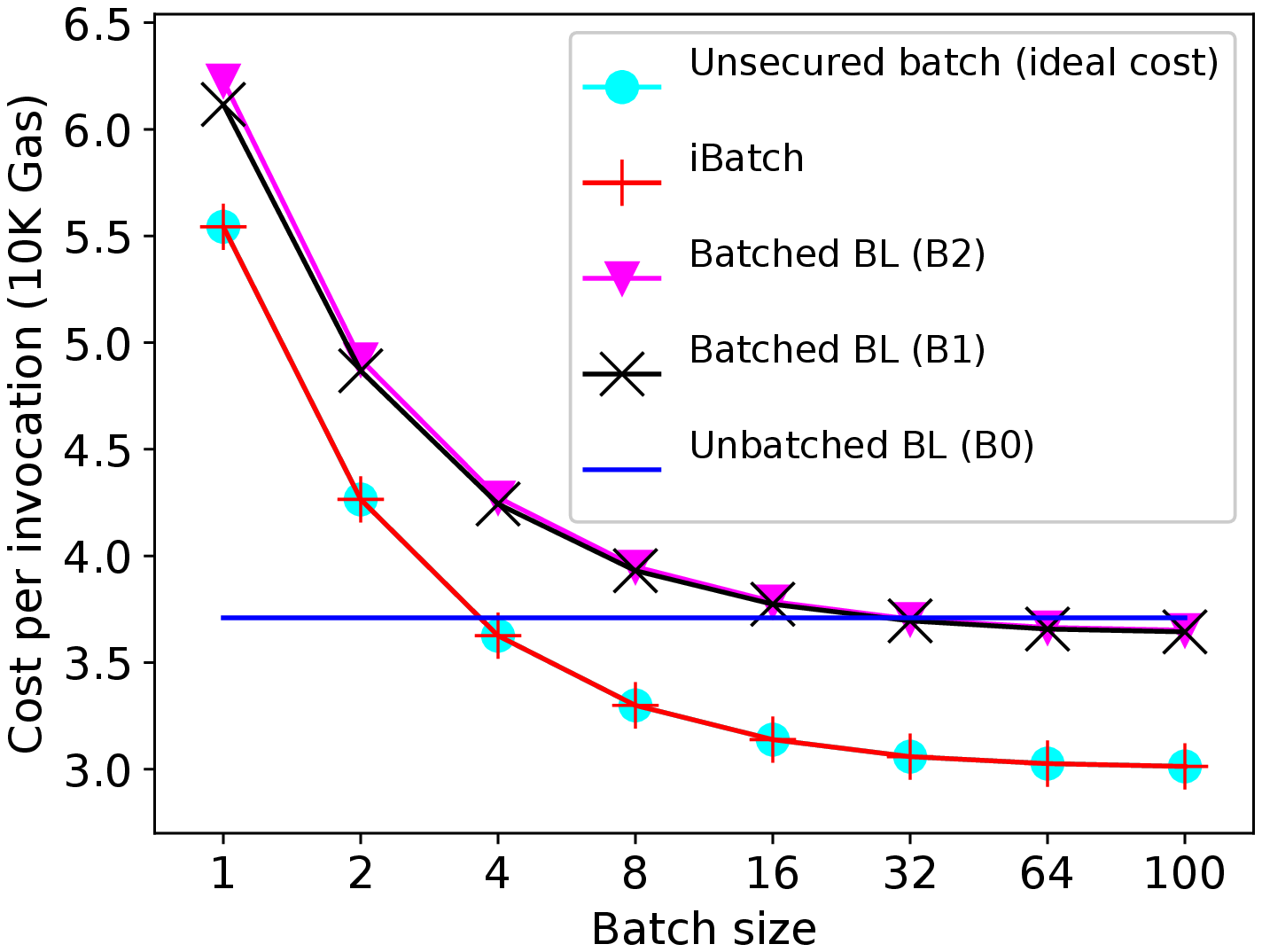}
    \label{fig:batchsize:1}
  }
  \subfloat[\sysname variants]{%
    \includegraphics[width=0.25\textwidth]{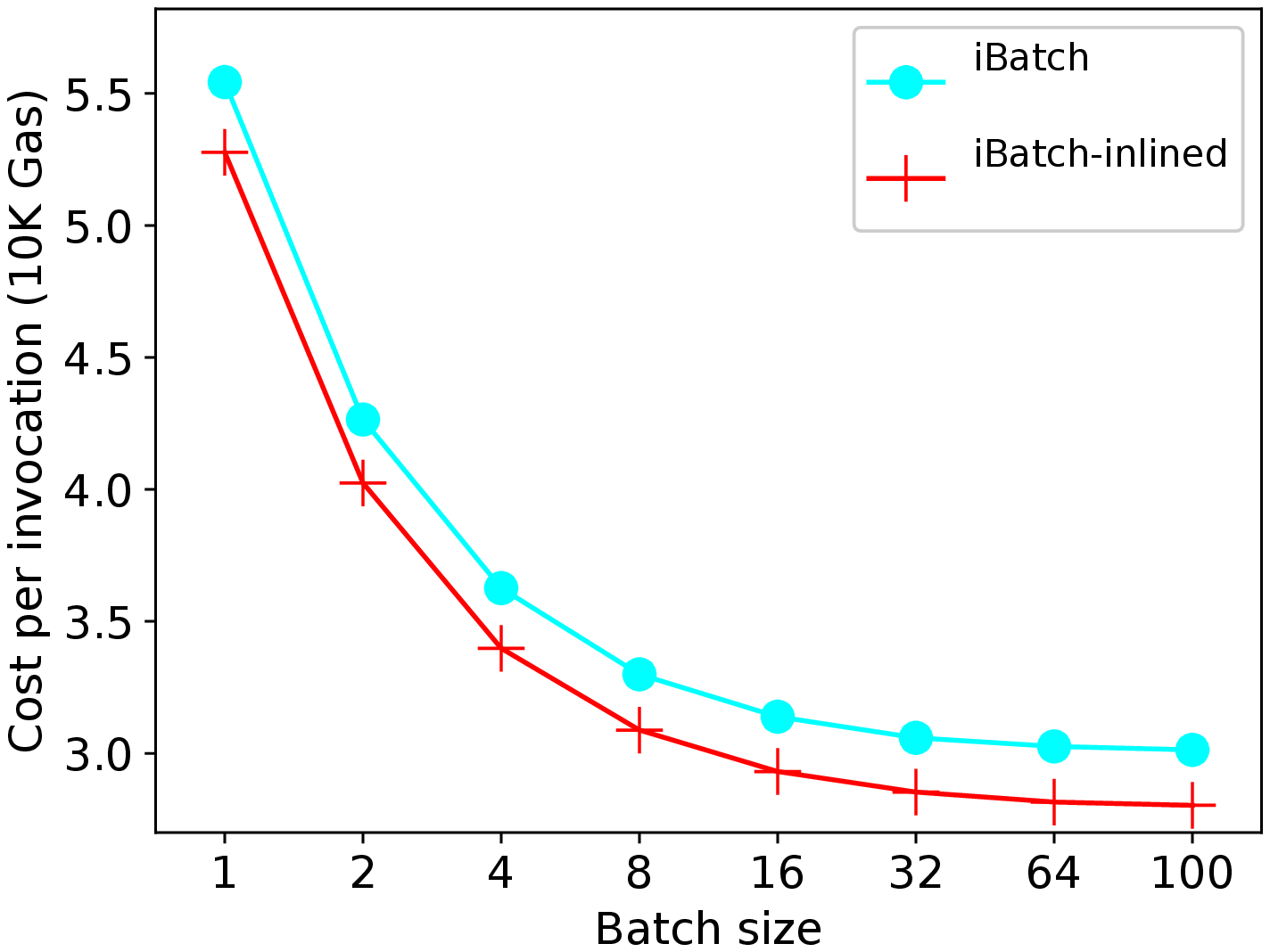}
    \label{fig:batchsize:2}
  }
\caption{Gas cost with varying batch sizes (BL refers to baselines)}
\end{figure}

%\begin{figure}
%\centering
%\includegraphics[width=0.3\textwidth]{figures/Parameter.eps}
%\caption{Gas cost with varying number of function arguments}
%\label{fig:argnumber}
%\end{figure}

This experiment aims to evaluate \sysname's cost in comparison with various baselines. 
We use a synthetic trace, that is, given $X$ batch size, we drive $X$ requests into target system (\sysname and other baselines) and measure their total Gas costs from the receipts of batch transactions. In the experiment, the target system is configured with BNB token contract~\cite{me:binance:contract}. In particular, we initialize the token balance of EOAs involved in the \texttt{transfer()} invocations, such that each \texttt{transfer()} invocation updates, instead of inserting, entries in the token balances. By this means, the Gas spent on executing \texttt{transfer()} is a constant. 
During the experiment, we vary the batch size $X$. In addition to \sysname and its inline variant, we measure the costs of various baselines, including the unbatched B0, batching \texttt{transfer} by \texttt{approve} of Dispatcher (B1),  batching with on-chain nonces (B2), and a cost-ideal approach of batching without defense against replay attacks (Ideal). Particularly, in the last approach, the \texttt{Dispatcher} does not maintain any nonce and verifies the callers' signatures, each of her own invocation. 

The results are reported in Figure~\ref{fig:batchsize:1}. It can be seen that baselines B1 and B2 have similar Gas cost, because on both cases the dispatcher contract need to maintain/update one word per invocation (i.e., allowance for B1 and on-chain nonce for B2). \sysname has similar costs with the batching ideal approach (without protection against replaying attacks), because both approaches do not maintain any states in smart contracts. 
With a sufficiently large batch (e.g., a batch of $100$ invocations), \sysname saves Gas by $14\%$ when compared with B1/B2, and by $17\%$ when compared with B0. When the batch size is smaller than $4$, \sysname can cause higher Gas than the baselines. Thus it only makes sense to have a batch containing more than $4$ invocations. 
This result is consistent with our cost analysis in \S~\ref{sec:costanalysis}. 
Besides, since the unbatched baseline (B0) is not batched, its cost is constant and irrelevant to the batch size.

Figure~\ref{fig:batchsize:2} presents the cost of \sysname and its inline variant with varying batch sizes. The inlined variant of \sysname saves $6.5\%$ Gas on top of \sysname, despite the batch size. 
}
{
\color{black} 
\section{Cost Analysis \& Deriving $N_{min}$}
\label{sec:costanalysis}
In this section, we conduct cost analysis of \sysname in comparison with the two baselines (B0 and B1). We use Ethereum's Gas-based cost model~\cite{wood2014ethereum} where selected operations and their costs are in Table~\ref{tab:costmodel}. One of the purposes is to derive a proper value of $N_{min}$, the minimal threshold of invocations in a batch.

\begin{table}[!htbp] %force in current page, disable float.
\caption{Ethereum's Gas cost model}
\label{tab:costmodel}\centering{\small
\begin{tabularx}{0.45\textwidth}{ |X|l| }
  \hline
 Operation & Gas cost ($X$ 32-byte words) \\ \hline
 Transaction & $G_{tx}(X) = 21000+2176X$ ($X<1000$) \\ \hline
 Internal call & $G_{call}(X) = 700+2176X$ ($X<1000$) \\ \hline
 Storage write (insert) & $G_{sset}(X)=20000X$ \\ \hline
 Storage write (update) & $G_{reset}(X)=5000X$ \\ \hline
 Storage read & $G_{sload}(X) = 200X$ \\ \hline
 Hash computation & $G_{sha3}(X) = 30+6X$ \\ \hline
\end{tabularx}
}
\end{table}

\begin{itemize}
\item {\bf B0}: In the unbatched baseline (B0), the $N$ requests will be sent in $N$ different transactions, resulting in the cost below. Here, $X$ is the request size (as stored in the data field of an Ethereum transaction) and $Y$ is the average contract cost per request.

\vspace{-0.15in}

{\scriptsize
\begin{eqnarray}
C_{B0}&= & 
(G_{tx}+G_{exec\_app})N
\label{eqn:cost:nobatch}
\\ \nonumber
& = &
(21000+2176X)N+YN
\\ \nonumber
& = &
21000*N+2176*X*N+Y*N
\end{eqnarray}
}

\item
{\bf B1}: 
The cost of the batching baseline (B1) is below, given that the $N$ requests are sent in one transaction. Note that $2176$ is the per-word cost of an internal call, $5000$ is the cost of verifying a $65$-byte signature, and the other $5000$ is the cost of writing a word on storage. $X+2$ is due to that both the account and signature are included in the transaction. 

\vspace{-0.15in}

{\scriptsize
\begin{eqnarray}
C_{B1}&=& 
G_{tx}(N)+(G_{exec\_\texttt{dispatch}}+G_{call}+G_{exec\_app})N
\label{eqn:cost:batchbaseline}
\\ \nonumber
& = &
(21000+2176(X+2)N)+(5000+5000)N
\\ \nonumber
&  &
+(700+2176X)N+YN
\\ \nonumber
& = &
21000+12877N+4352XN+YN
\end{eqnarray}
}

\item
{\bf\sysname}: 
The cost of \sysname is below, given that the $N$ requests are sent in one transaction. 

\vspace{-0.15in}
{\scriptsize
\begin{eqnarray}
C_{B1}&=& 
G_{tx}(N)+(G_{exec\_\texttt{dispatch}}+G_{call}+G_{exec\_app})N
\label{eqn:cost:etherbatch}
\\ \nonumber
& = &
(21000+2176(X+2)N)+(5000)N
\\ \nonumber
& &
+(700+2176(X+1))N+Y'N
\\ \nonumber
& \approx &
21000+10053N+4352XN+Y'N
\end{eqnarray}
}
In the last step, we assume the rewritten contract has a similar Gas cost with the original contract (i.e., $Y'\approx{}Y$).
\end{itemize}

Comparing Equation~\ref{eqn:cost:nobatch} and Equation~\ref{eqn:cost:batchbaseline} (i.e., the costs between B0 and B1), we have: 

\vspace{-0.15in}
{\scriptsize
\begin{eqnarray}
\frac{C_{B0}}{C_{B1}} - 1 &=& \frac{21000N+2176NX+YN}{21000 + 12877N + 4352XN+YN} - 1 
\\ \nonumber
&=& \frac{8123 - 2176X - 21000/N}{21000/N + 12877 + 4352X+Y}
\end{eqnarray}
}

With a common setting $X=3$, this leaves the saving by batching B1 quite trivial. As will be seen in real experiments, B1 actually increases the Gas instead of saving (\S~\ref{appdx:sec:eval:micro}).

Comparing Equation~\ref{eqn:cost:nobatch} and Equation~\ref{eqn:cost:etherbatch} (i.e., the costs between B0 and \sysname), we have: 

\vspace{-0.15in}
{\scriptsize
\begin{eqnarray}
\frac{C_{B0}}{C_{\sysname}} - 1 &=& \frac{21000N+2176NX+YN}{21000 + 10053N + 4352XN+YN} - 1 
\\ \nonumber
&=& \frac{10947 - 2176X - 21000/N}{21000/N + 10053 + 4352X+Y}
\end{eqnarray}
}

With $X=3$, if \sysname has a lower Gas cost than non-batching baseline B0, it requires: 

\vspace{-0.15in}
{\scriptsize
\begin{eqnarray}
&&10947 - 2176*3 - 21000/N > 0
\\ \nonumber
&\Rightarrow& N > 21000/4419 = 4.75\defeq{N_{min}}
\end{eqnarray}
}
In our experiment, we set $N_{min}$ to be $5$, that is, only when a time interval containing more than $5$ calls will lead to a batched transaction.

Comparing Equation~\ref{eqn:cost:batchbaseline} and Equation~\ref{eqn:cost:etherbatch}, it is clear that the \sysname can save more Gas than the batching baseline (without modifying contracts). This is because in the baseline, the \texttt{Dispatcher} contract needs to write a storage state upon each internal call (e.g., \texttt{allowance} when using the \texttt{approve/transferFrom()} workflow).
}
{
\color{black} 
\section{Discussion}
\subsection{Batching Ether Transfers}
\label{sec:paymentbatching}

The idea of batching can not only be applied to smart contract invocations but also Ether payments. Without batching, each Ether payment costs the fee of one transaction, that is, $G_{tx} = 21000$ Gas. 

Batching Ether payments in one transaction works in a similar way with \sysname: The \texttt{PaymentDispatcher} contract initially receives Ether deposits from an owner and then, upon the owner's request, transfers the Ether on behalf of her. In each request, the parameters of the Ether payment are embedded in the function arguments of a \texttt{PaymentDispatcher} contract. The \texttt{PaymentDispatcher} contract verifies the parameters against the owner's public key (blockchain address). This is similar to \sysname's \texttt{Dispatcher} contract with one difference: Instead of issuing an internal call, \texttt{PaymentDispatcher} issues a \texttt{transfer()} call~\cite{me:ether:transfer}.

The cost of the batching Ether payment scheme above is the following: Given $N$ payments batched in one transaction, the per-payment Gas is as following. Here, the signature is of 65 bytes. Both addresses (\texttt{from} and \texttt{to}) are of $20$ bytes.
Gas $5000$ is for verifying the signature and $7800$ is the cost of an internal call to \texttt{transfer()}.

\begin{eqnarray}
\nonumber
&&
\frac{21000}{N}+2176*(65/32+20/32+20/32+1)
\\\nonumber
&&
+5000+7800
\\\nonumber
&=& 21000\frac{1}{N} + 22116 > 21000
\end{eqnarray}

{\color{black}
\subsection{\sysname Beyond Ethereum}

We first conduct a generic analysis to derive a necessary condition to make \sysname profitable, and then examine this conduction for real-world blockchain platforms.

{\bf \sysname's profitable condition}: A generic transaction consists of two parts: 1) A minimal transaction of 3 words to transfer coins (3 words for sender address, receiver address and the amount of ``coins'' transferred) and2) the data field of $N$ words necessary for triggering smart contract execution. Thus, a transaction is of $3+N$ words. We here consider a linear cost model where a transaction's fee is modeled as $X+N*Y$ where $X$ denotes the cost of a transfer-only transaction ($3$-word long), and $Y$ is the unit cost per word in the data field.

Consider two invocations, each of $M$ arguments. We represent each invocation by a $4+M$-word triplet (recall Equation~\ref{eqn:1:invocation}).  Placing two invocations in one transaction would lead to transaction fee being $X + 2*(4+M)*Y$. Placing two invocations in two separate transaction has a fee of  $2[X+(2+M)Y]$ (because the caller address and callee contract address can be encoded by the native sender and receiver of the transfer-only part of the transaction, leaving the data field of the transaction to be 4+M-2=2+M words). 

Thus, the net saving of transaction fee by \sysname is $2[X+(2+M)Y]-[X+2(4+M)Y]=X-4Y$. In other words, to make \sysname result in positive net fee saving, it entails to check the following inequality:

\begin{equation}
X-4Y\stackrel[]{?}{>}0
\end{equation}

{\bf Case studies}: In Ethereum, $X=21000$ and $Y=2176$ (recall Table~\ref{tab:costmodel}). Thus, it holds $X-4Y=21000-4*2176>0$.

{\it The case of TRON}: 
The tron blockchain~\cite{me:tron} is similar to Ethereum in that it supports smart contracts written in Turing complete languages and charges smart contract execution by Gas like cost. 

The TRON blockchain has two cost metrics, ``Bandwidth'' and ``Energy''. The Energy cost applies only to smart-contact execution, thus for transaction fee saving, we consider only TRON's Bandwidth cost. By accessing TRON's shasta testnet~\cite{me:tron:shasta}, we derive TRON's cost model that $X=267$ and $Y=47$. Thus, $X-4Y=267-4*47=79>0$, which implies the applicability/profitability of \sysname approach to the TRON blockchain.

\begin{table}[!htbp] %force in current page, disable float.
\caption{Cost before/after Batching in EOS}
\label{tab:eos:batching}\centering{\small
\begin{tabularx}{0.45\textwidth}{ |X|p{1.5cm}|p{1cm}| }
  \hline
   & ``Bandwidth'' per call & ``CPU'' per call \\ \hline
   Two calls in two transactions & 104 & 228 \\ \hline
   Two calls in two actions in one transaction & 72 & 153  \\ \hline
   Two calls batched in one action & 57 & 156\\ \hline
\end{tabularx}
}
\end{table}

{\it The case of EOS}: 
EOS~\cite{me:eosio} is a popular blockchain supporting expressive smart contracts. It charges transaction fee and contract execution in three metrics, ``CPU'', ``Bandwidth'', and ``RAM'' storage. Sending transactions without causing smart contract execution does not cost RAM. Also, in EOS, the CPU cost of a transaction is not linear w.r.t. the transaction length (i.e., it does not match our cost model here), and we leave it to our empirical study. So here, we only consider Bandwidth. For EOS bandwidth, $X=128$ and $Y=8$. Thus, $X-4Y=96>0$.

We also run an EOS.IO node locally~\cite{me:eos:localnode} and conducted measurement study on EOS's CPU and Bandwidth. In this study, we consider two invocations to a simple helloworld smart contract~\cite{me:eos:helloworld}. When putting them into two separate transactions, the CPU cost per call is on average 228 ``usec'' (Note that the CPU cost is dependent on runtime/hardware and is non-deterministic). The Bandwidth cost per call is 104 bytes. When putting the two calls in two actions of one transaction, the Bandwidth cost per call is 144/2=72 bytes and CPU is about 305/2=153 usec. When putting the two calls in one action, the CPU cost is 311/2=156 usec and Bandwidth cost is 114/2=57 bytes.

Note that EOS adopts a ``Receiver Pay'' model (that is, contract execution cost is charged to contract creator's account, not transaction sender's account), and the saving by \sysname applies to the contract creator as well.
}
}

{
\color{black} 
\section{Backward Compatibility}
\label{sec:compatible} 
For backward-compatibility, the rewritten function should preserve the ``functionality'' of the original function and have the same effects on the blockchain state. Intuitively, an owner $o$ invoking the original function \texttt{foo} is equivalent with the \texttt{Dispatcher} contract invoking the rewritten function \texttt{fooByD} on behalf of owner $o$. Formally, we consider a stateful-computation model for contract execution and describe the backward-compatibility below:

\begin{definition}[Backward-compatibility]
Suppose owner $o$ invokes a smart-contract function \texttt{foo} with arguments $args$. The function invocation returns $output$. The invocation also transitions the contract state from initial state $st$ to end state $st'$. We denote the contract invocation by  $(st', output) = o.\texttt{foo}(st, args)$.

Rewritten function \texttt{fooByD(from, $args$)} is backward-compatible or functional-equivalent with original function \texttt{foo($args$)}, if and only if for any invocation $(st', output) = o.\texttt{foo}(st, args)$, we have $(st', output) =\texttt{Dispatcher.fooByD}(st, o, args)$.
\end{definition} 

Backward-compatibility implies that one can freely replace any contract function with its counterpart in any context, that is, without affecting other function execution (i.e., replacing \texttt{foo} with \texttt{fooByD} or replacing \texttt{fooByD} with \texttt{foo}). For instance, an HTLC created by \texttt{newContractByD} can be withdrawn by the original receiver calling \texttt{withdraw()}. Likewise, an HTLC created by \texttt{newContract} can be refunded by the \texttt{Dispatcher} invoking \texttt{refundByD()} on behalf of the original sender.

{\bf Evaluation}: We verify that the \sysname design is functional by writing a series of test programs. Each test program issues a sequence of function calls to a specific application contract. For the ERC20 token, the test program issues \texttt{transfer} calls. For the IDEX, the test program issues \texttt{deposit} and \texttt{trade} calls. For the HTLC, the test program covers two cases: 1) \texttt{newContract} and \texttt{withdraw} and 2) \texttt{newContract} and \texttt{refund}. 

We verify that the \sysname design is functional by writing a series of test programs. Each test program issues a sequence of function calls to a specific application contract. For the ERC20 token, the test program issues \texttt{transfer} calls. For the IDEX, the test program issues \texttt{deposit} and \texttt{trade} calls. For the HTLC, the test program covers two cases: 1) \texttt{newContract} and \texttt{withdraw} and 2) \texttt{newContract} and \texttt{refund}. 

We then ``fuzz'' the smart-contract invocations in each of these test programs. That is, each invocation can be carried out by a dedicated transaction (as an external call) or by a batched transaction (in \sysname). Since each test program contains less than $N=4$ function calls, there are at most $2^N=16$ possible call combinations after the ``fuzzing''. The results verify the "backward compatibility" of the rewritten contracts. That is, serving the \sysname's internal call to the rewritten contract has the same on-chain effects as serving the external call to the original application contract.

}

\section{Related Works}
\label{sec:rw}

Public blockchains are known to cause high costs and to have limited transaction throughput~\cite{DBLP:conf/fc/CromanDEGJKMSSS16}. Reducing the cost of blockchain applications is crucial for real-world adoption and has been studied in the existing literature. 
{
\color{black} 
{\bf Layer-one protocols}: 
Newer blockchains or so-called layer-one protocols are proposed such as sharding and other designs~\cite{DBLP:conf/sp/Kokoris-KogiasJ18,DBLP:conf/ccs/LuuNZBGS16}. Deploying these mechanism requires launching a new blockchain network from scratch, and it is known to be difficult to bootstrap a large-scale blockchain.
}
{\bf Layer-two protocols}: 
Another approach, dubbed layer-two designs~\cite{me:lightning,DBLP:journals/corr/MillerBKM17,DBLP:journals/corr/abs-1804-05141,DBLP:conf/ccs/DziembowskiFH18}, focuses on designing add-on to a deployed blockchain system by designing extensions including smart-contracts on-chain and services off-chain. The notable example is payment networks~\cite{me:lightning,DBLP:journals/corr/MillerBKM17,DBLP:journals/corr/abs-1804-05141} that place most application logic of making a series of micro-payments off the blockchain while resorting to blockchain for control operations (e.g., opening and closing a channel) and error handling. 
In a sense, a payment channel ``batches'' multiple repeated micro-payments into minimally two transactions.
State channels~\cite{DBLP:conf/ccs/DziembowskiFH18} generalize the idea to support the game-based execution of smart contracts.
The batching in this line of work is orthogonal to that in \sysname: 1) \sysname{} is generally applicable to any smart contracts, while payment channel/network is specific to repeated micro-payments between a fixed pair of buyer and seller. 2) \sysname{} can further reduce the Gas of a payment channel. Specifically, the invocations to the smart contracts in a payment channel (namely HTLC) can be batched to amortize the transaction fee over multiple operations to open/close a channel~\cite{me:lightning}. 
{\bf Ethereum Gas optimization}: 
GRuB~\cite{DBLP:journals/corr/abs-1911-04078} supports gas-efficient data feeds onto blockchains, for decentralized financial applications (DeFi). For Gas efficiency, it employs a novel technique that replicates data feeds adaptively to the workload. \sysname{} can complement GRuB's adaptive data-feed to achieve higher level of Gas efficiency.
Gasper~\cite{DBLP:conf/wcre/ChenLLZ17} detects and fixes the ``anti-patterns'' in smart contracts that excessively cost Gas. While Gasper aims at reducing the on-chain computation in smart contracts, \sysname{} reduces the transaction fee in smart-contract invocations.
{
\color{black} 
{\bf Transaction mixing}: 
The purpose of mixing~\cite{DBLP:conf/fc/BonneauNMCKF14,DBLP:conf/ndss/HeilmanABSG17,DBLP:journals/popets/MeiklejohnM18,DBLP:journals/iacr/SeresNBB19} is to hide the linkage between transaction senders and receivers. In Bitcoin where a transaction natively supports multiple coin transfers, mixing can be enabled by coordinating multiple senders to jointly generate a multi-input transaction with the input-output mapping shuffled (such transaction is called CoinJoin transaction); generating a CoinJoin transaction can be done without a trusted third-party~\cite{DBLP:conf/fc/BonneauNMCKF14,DBLP:conf/ndss/HeilmanABSG17,DBLP:journals/popets/MeiklejohnM18}. Mixing in Ethereum~\cite{DBLP:journals/iacr/SeresNBB19,me:ethmixer} works by having multiple senders send their coins to a mixer account who then sends the coins to original receivers. The process results in transactions twice the number of senders and does not lead to saving of transaction fee as \sysname does. In addition, note that mixing needs to break the linkage between senders and receivers, while batching needs to preserve such caller-callee linkage so that the callee smart contract can recognize and admit the caller.
{\bf Batching Ethereum transactions}:
A recent work batches ERC20 token invocations in the application of token airdropping~\cite{DBLP:conf/esorics/FrowisB19}. While the work solves a special case of batching (of a single callee function, i.e., \texttt{transfer}), \sysname is a more comprehensive and systematic work in the sense that it covers the general case of batching with multiple callers/callees and addresses the practical integration of batching with a deployed Ethereum platform.
OpenZeppelin's gas station network~\cite{me:openzepplin:gsn,me:openzepplin:gsncontracts} supports invocations from users without Ether wallets by similarly extending Ethereum with on-chain/off-chain components, but does not particularly address batching.
}

\section{Conclusion}
\label{sec:conclude}
This paper presents \sysname, a security protocol and middleware system to batch smart-contract invocations over Ethereum. The design of \sysname{} addresses the tradeoff between security, cost effectiveness and delay. 
The result shows that compared with the baseline without batching, \sysname{} effectively saves cost per invocation with small block delay. 

\section*{Acknowledgments}
The authors appreciate the anonymous reviewers and shepherd.
The first five authors at Syracuse University are partially supported by the National Science Foundation under Grant CNS1815814 and DGE2104532.
Xiapu Luo is partially supported by Hong Kong RGC Project 152223/20E and Hong Kong ITF Project GHP/052/19SZ.
Ting Chen is partially supported by Project 2018YFB0804100 under National Key R\&D Program of China and Project 61872057 under National Natural Science Foundation of China.

\bibliographystyle{abbrv}
{
\bibliography{main}
}

%\appendix
%\input{text/wellformated_appendix.tex}
%\clearpage\vspace{0.2in}\noindent{}{\LARGE \bf Supplementary Material}
%\input{text/v1_evaluateEther.tex}
%\input{text/techrep_appendix.tex}
%\clearpage\input{text/6_eval_macro.tex}

\end{document}